\documentclass[
floatfix,
twocolumn,
showpacs,
preprintnumbers,
amsmath,amssymb,pra,superscriptaddress,longbibliography]{revtex4-2}

\usepackage{color}
\usepackage[usenames,dvipsnames,svgnames,table]{xcolor}
\usepackage[colorlinks=true,linkcolor=blue,urlcolor=blue,citecolor=blue]{hyperref}
\usepackage{graphicx}

\usepackage{bmpsize}
\usepackage{dcolumn}

\usepackage{bm}
\usepackage{xcolor}
\usepackage{enumerate}
\usepackage{booktabs}
\usepackage{threeparttable}

\newcommand{\br}{\bm{r}}
\newcommand{\bR}{\bm{R}}
\newcommand{\ubr}{\underline{\bm{r}}}
\newcommand{\ubR}{\underline{\bm{R}}}
\newcommand{\bu}{{\bf u}}
\newcommand{\bU}{{\bf U}}
\newcommand{\bF}{{\bf F}}
\newcommand{\s}{_\mathrm{{\scriptscriptstyle S}}}

\newcommand{\xc}{_\mathrm{{\scriptscriptstyle XC}}}

\newcommand{\bk}{\bm{k}}

\newcommand{\yellow}[1]{\noindent{{\color{yellow!60!black} #1}}}
\newcommand{\green}[1]{\noindent{{\color{green!60!black} #1}}}
\newcommand{\blue}[1]{\noindent{{\color{blue!60!black}  #1}}}

\begin{document}


\title{Accelerating Finite-Temperature Kohn-Sham Density Functional Theory\\ 
with Deep Neural Networks}

\author{J.A. Ellis}
\affiliation{Sandia National Laboratories, Albuquerque, NM 87185, USA}

\author{L. Fiedler}
\affiliation{Center for Advanced Systems Understanding (CASUS), D-02826 G\"orlitz, Germany}
\affiliation{Helmholtz-Zentrum Dresden-Rossendorf, D-01328 Dresden, Germany}

\author{G.A. Popoola}
\affiliation{Sandia National Laboratories, Albuquerque, NM 87185, USA}

\author{N.A. Modine}
\affiliation{Sandia National Laboratories, Albuquerque, NM 87185, USA}

\author{J.A. Stephens}
\affiliation{Sandia National Laboratories, Albuquerque, NM 87185, USA}

\author{A.P. Thompson}
\affiliation{Sandia National Laboratories, Albuquerque, NM 87185, USA}

\author{A. Cangi}
\email{a.cangi@hzdr.de}
\affiliation{Center for Advanced Systems Understanding (CASUS), D-02826 G\"orlitz, Germany}
\affiliation{Helmholtz-Zentrum Dresden-Rossendorf, D-01328 Dresden, Germany}

\author{S. Rajamanickam}
\email{srajama@sandia.gov}
\affiliation{Sandia National Laboratories, Albuquerque, NM 87185, USA}

\date{\today}

\begin{abstract}
We present a numerical modeling workflow based on machine learning (ML) which reproduces the total energies produced by Kohn-Sham density functional theory (DFT) at finite electronic temperature to within chemical accuracy at negligible computational cost. 
Based on deep neural networks, our workflow yields the local density of states (LDOS) for a given atomic configuration. From the LDOS, spatially-resolved, energy-resolved, and integrated quantities can be calculated, including the DFT total free energy, which serves as the Born-Oppenheimer potential energy surface for the atoms. 
We demonstrate the efficacy of this approach for both solid and liquid metals and compare results between independent and unified machine-learning models for solid and liquid aluminum.
Our machine-learning density functional theory framework opens up the path towards multiscale materials modeling for matter under ambient and extreme conditions at a computational scale and cost that is unattainable with current algorithms. 
\end{abstract}

\maketitle



\section{\label{sec:introduction} Introduction}

Multiscale materials modeling~\cite{Hor2012:future} provides fundamental insights into microscopic mechanisms that determine materials properties. A multiscale modeling framework operating both near first-principles accuracy and across length and time scales would enable key progress in a plethora of applications. It would greatly advance materials science research---in gaining understanding of the dynamical processes inherent to advanced manufacturing~\cite{Fra2014:metal, AD2018:advanced}, in the search for superhard~\cite{MMET2003:bonding} and energetic materials~\cite{FMPS2001:design, PLMS2002:review}, and in the identification of radiation-hardened semiconductors~\cite{WF2001:review, Was2007:fundamentals}. It would also pave the way for generating more accurate models of materials at extreme conditions including shock-compressed materials~\cite{GMVC1986:materials,Saw1986:role} and astrophysical objects, such as the structure, dynamics, and formation processes of the Earth’s core~\cite{AG98}, solar~\cite{Militzer_2008,militzer1,manuel, NBHR12,NHFR13, LHR09,LHR11} and extra-solar planets\cite{NFKR11,KKNF12}.
This requires faithfully passing information from quantum mechanical calculations of electronic properties on the atomic scale (nanometers and femtoseconds) up to effective continuum material simulations operating at much larger length and time scales. Atomistic simulations based on molecular dynamics (MD) techniques~\cite{AW1959:studies} and their efficient implementation~\cite{Pli1995:fast} for large-scale simulations on high-performance computing platforms are the key link. 
For atomic-scale systems, MD results can be directly validated against quantum calculations.  At the same time, MD can be scaled up to the much larger scales on which the continuum behaviors start to emerge (either micrometers or microseconds, or even both~\cite{Zepedaruiz2017}).

MD simulations require accurate interatomic potentials (IAPs)~\cite{Fin2003:interatomic}. Generating them based on machine learning (ML) techniques has only recently become a rapidly evolving research area. It has already led to several ML-generated IAPs such as  
AGNI~\cite{HBC+2017:universal},
ANI~\cite{Smith2017},
DPMD~\cite{ZHW+2018:deep},
GAP~\cite{BPKC2010:gaussian}, 
GARFfield~\cite{JNG2014:general},
HIP-NN~\cite{LSB2018:hierarchical}, SchNet~\cite{SSK+2018:schnet}, 
and SNAP~\cite{TST+2015:spectral}. While these IAPs differ in their flavor of ML models, for instance, non-linear optimization, kernel methods, structured neural networks, and convolutional neural networks, they all rely on accurate training data from first-principles methods.

First-principles training datasets are commonly generated with Kohn-Sham density functional theory (DFT)~\cite{KS1965:selfconsistent} and its generalization to finite electronic temperature~\cite{Mer1965:thermal}. Due to its balance between accuracy and computational cost, it is the method of choice~\cite{Bur2012:perspective} for computing properties of molecules and materials with close to chemical accuracy~\cite{Bec2014:perspective,LBB+2016:reproducibility}. 

Despite its success~\cite{PGB2015:dft}, DFT is limited to the nanoscale due to its computational cost which scales as the cube of the system size. At the same time, the amount of data needed to construct an IAP increases exponentially with the number of chemical elements, thermodynamic states, phases, and interfaces~\cite{WCWT2019:datadriven}. Likewise, evaluation of thermodynamic properties using DFT is computationally expensive, typically $10^5$ to $10^6$ core-hours for each point on a grid of temperatures and densities~\cite{WCR+2018:shock}. Furthermore, fully converged DFT calculations at low densities, very high temperatures, or near phase transitions are considerably more difficult.

\begin{figure*}[htpb]
	\centering
	\includegraphics[width=1.9\columnwidth]{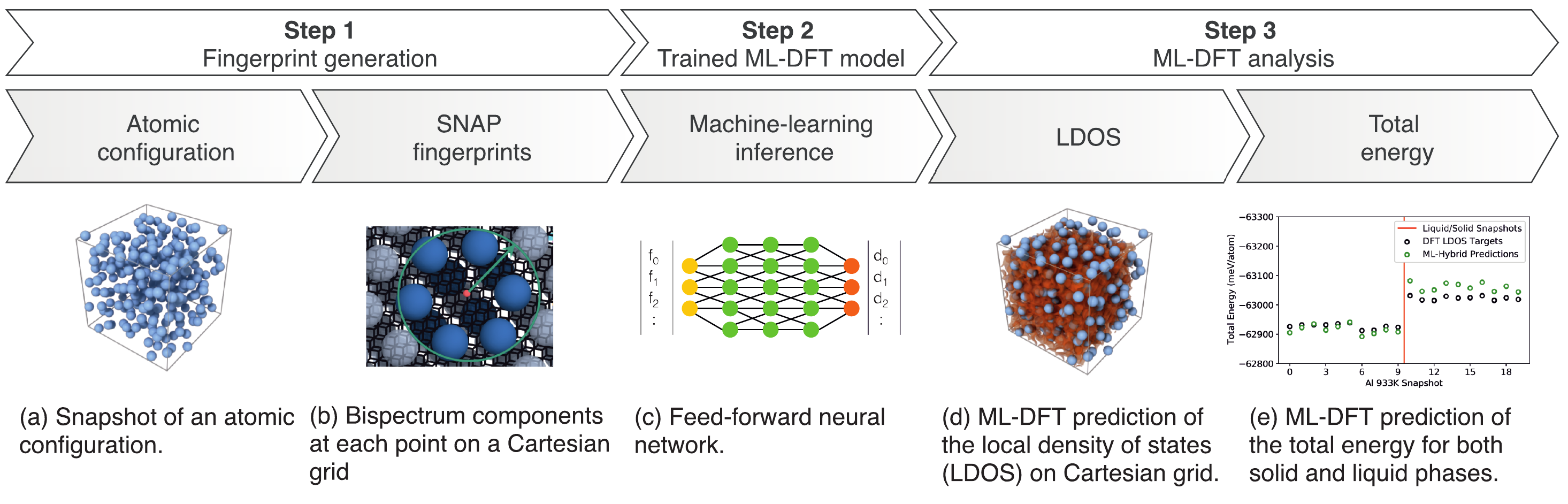}
	\caption{Illustration of the ML-DFT workflow listed in Table~\ref{tab:workflow}. 
		The raw input data is generated from finite-temperature DFT calculations for snapshots of atomic configurations, i.e., positions and chemical identities of atoms (a). SNAP fingerprints in terms of bispectrum components are generated for each of several million Cartesian grid points in the snapshot volume (b). The trained ML-DFT feed-forward neural network takes the input vector $\boldsymbol{f}$ composed of SNAP fingerprints for one Cartesian grid point and outputs  the predicted LDOS vector $\boldsymbol{d}$ at the grid point (c). The primary output of the ML inference is the LDOS on a Cartesian grid (d) which is used to compute the total energy (e) based on Eq.~(\ref{eq:FES.LDOS}). The embedded Figure~\ref{fig:hybrid_bes} displays our central result---the capability of a single ML-DFT model to predict the total energy in both solid and liquid phases of aluminum with accuracy comparable to standard DFT.}
	\label{fig:ml.workflow}
\end{figure*}

Pioneering efforts in applying ML to electronic structure calculations focus on mining benchmark databases generated from experiments, quantum chemistry methods, and DFT in order to train ML models based on kernel ridge regression and feed-forward neural networks. These efforts enabled ML predictions of molecular properties~\cite{HBR+2015:machine,RTMv2012:fast,SAC+2017:quantumchemical} and crystal structures~\cite{SGB+2014:how}. Only limited efforts have gone into actually speeding up DFT calculations by directly approximating the solutions of the Kohn-Sham differential equations through ML models. Prior work has focused on approximating the kinetic energy functional using ML regression~\cite{LSP+2016:understanding,SRH+2013:orbitalfree,SRH+2012:finding} and deep-learning models~\cite{YP2016:kinetic}. Most recently, ongoing efforts have demonstrated the usefulness of ML kernel techniques on the electronic density~\cite{BVL+2017:bypassing} and feed-forward neural networks on the density of states (DOS)~\cite{Broderick2011, SGB+2014:how, mahmoud2020learning}, and the local density of states (LDOS)~\cite{CKB+2019:solving} for creating computationally efficient DFT surrogates. 

In this work, we develop ML-DFT, a surrogate for DFT calculations at finite electronic temperature using feed-forward neural networks. We derive a representation of the Born-Oppenheimer potential energy surface at finite electronic temperature (Section~\ref{ssec:bo-fes}) that lends itself to an implementation within an ML workflow (Section~\ref{ssec:ml-fes}). 
Using the LDOS as the central quantity, ML-DFT enables us to compute the DFT total free energy in terms of a single ML model (Section~\ref{ssec:ml-fes}), applicable to coupled electrons and ions not only under ambient conditions, but also in liquid metals and plasmas. For the sake of simplicity, we will refer to the DFT total free energy as defined in Eq.~(\ref{eq:fes}) as total energy.
Utilizing a generalization of the SNAP bispectrum descriptors on a Cartesian grid (Section~\ref{ssec:fingerprint}), the feed-forward neural network of ML-DFT (Section~\ref{ssec:ml_workflow}) is trained on DFT data (Section~\ref{ssec:dft-data}).  Integrated quantities such as the band energy and the total energy of an atomic configuration at a given electronic temperature are in turn computed based on the LDOS predicted by ML-DFT. Using the LDOS in conjunction with the developed expression for total energy avoids the need to introduce another ML model to map densities to kinetic energies~\cite{BVL+2017:bypassing} and goes significantly beyond prior ML efforts related to the LDOS~\cite{CKB+2019:solving}. 
ML-DFT yields DFT-level accuracy, but comes at a negligible computational cost by avoiding the $O(N^3)$ eigensolver step. 
We demonstrate this for aluminum (Section~\ref{sec:results}). Trained on atomic configurations of aluminum at the melting point, a single ML-DFT model yields accurate results (LDOS, electronic density, band energy, and total energy) for both crystalline and liquid aluminum.  Notably, band energies and total energies predicted by ML-DFT agree with the results of conventional DFT calculations to well within \emph{chemical accuracy}, which is traditionally defined as 1 kcal/mol = 43.4 meV/atom. Hence, ML-DFT meets all requirements to become the backbone of the computational infrastructure needed for high-performance multiscale materials modeling of matter under ambient and extreme conditions (Section~\ref{sec:conclusion}).

The ML-DFT workflow is illustrated in Figure~\ref{fig:ml.workflow}. It starts with input data generated from finite-temperature DFT calculations for a given atomic configuration (a). SNAP fingerprints are generated for each of several million Cartesian grid points in the snapshot volume (b). Given a SNAP fingerprint as input, the trained ML-DFT feed-forward neural network makes its prediction for the LDOS vector (c). The primary output of the ML inference is the LDOS prediction at each grid point (d). Based on the predicted LDOS, quantities such as the electronic density based on Eq.~(\ref{eq:n.from.ldos}), the DOS based on Eq.~(\ref{eq:dos.from.ldos}), and the total energy based on Eq.~(\ref{eq:FES.LDOS}) are computed (e).

\section{\label{sec:methods} Methods}

\subsection{\label{ssec:bo-fes} Born-Oppenheimer Density Functional Theory at Finite Electronic Temperature}

A suitable theoretical framework for computing thermodynamic materials properties from first principles is within the scope of non-relativistic quantum mechanics in the Born-Oppenheimer approximation~\cite{AMG2012:correlated}. Additionally we work within a formalism that takes the electronic temperature into account. This becomes relevant when the scale of the electronic temperature becomes comparable to the first excitation energy of the electronic system which is particularly relevant for liquid metals and plasmas. In the given framework we, hence, assume that the electrons find their thermal equilibrium on a time scale that is small compared to that of ionic motion. The formal development and implementation of such methodologies that couple electron and ion dynamics is an area of active research~\cite{GHMC1994:initio,  AKPF1994:initio, APF1995:initio,  MVP1997:ensemble,ACE+2010:initio, PhysRevLett.105.123002,SMS2016:exploring}.

More formally, consider a system of $N_e$ electrons and $N_i$ ions with collective coordinates $\ubr=\{\br_1,\dots,\br_{N_e}\}$ and $\ubR=\{\bR_1,\dots,\bR_{N_i}\}$, where $\br_j \in \mathbb{R}^3$ refers to the position of the $j$th electron, while $\bR_\alpha \in \mathbb{R}^3$ denotes the position of the $\alpha$th ion of mass $M_\alpha$ and charge $Z_\alpha$. 
The physics in this framework is governed by the Born-Oppenheimer Hamiltonian
\begin{align}
	\label{eq.bo-hamiltonian}
	\hat{H}^{BO}(\ubr; \ubR) &= \hat{T}^{e}(\ubr) + \hat{V}^{ee}(\ubr) + \hat{V}^{ei}(\ubr; \ubR) + \hat{V}^{ii}(\ubR)
\end{align}
where $\hat{T}^{e}(\ubr) = \sum_{j}^{N_e} -\nabla_j^2/2$ denotes the kinetic energy of the electrons, $\hat{V}^{ee}(\ubr) = \sum_{j}^{N_e} \sum_{k\neq j}^{N_e} |\br_j -\br_k|^{-1}/2$ the electron-electron interaction, $\hat{V}^{ei}(\ubr; \ubR) = -\sum_{j}^{N_e} \sum_{\alpha}^{N_i} Z_\alpha |\br_j-\bR_\alpha|^{-1}$ the electron-ion interaction, and $\hat{V}^{ii}(\ubR)=\sum_{\alpha}^{N_i} \sum_{\beta \neq \alpha}^{N_i} Z_\alpha Z_\beta |\bR_\alpha -\bR_\beta|^{-1}/2$ the ion-ion interaction. Note that we work within atomic units throughout, where $\hbar = m_e = e^2 = 1$, such that energies are expressed in Hartrees and length in Bohr radii.

Using the standard framework of quantum statistical mechanics~\cite{TKS1983:statistical}, all thermodynamic equilibrium properties are formally given in terms of the grand potential
\begin{align}
	\label{eq.def.gp}
	\Omega[\hat{\Gamma}]&=\mathrm{Tr}\left(\hat{\Gamma} \hat \Omega\right)
\end{align}
which is defined as a statistical average over the grand canonical operator $\hat \Omega = \hat{H}^{BO} - \mu \hat{N} - \hat{S}/\beta$, where $\hat{N}$ denotes the particle number operator, $\hat{S}$ the entropy operator, $\mu$ the chemical potential, and $1/\beta = k_B T$, where $k_B$ is the Boltzmann constant and $T$ is the electronic temperature.
Statistical averages as in Eq.~(\ref{eq.def.gp}) are computed via the statistical density operator $\hat{\Gamma} = \sum_{N_e,j} w_{N_e,j} |\Psi_{N_e,j}\rangle \langle \Psi_{N_e,j}|$ which is a weighted sum of projection operators on the underlying Hilbert space spanned by the $N_e$-particle eigenstates $\Psi_{N_e,j}$ of $\hat{H}^{BO}$ with energies $E_{N_e,j}$ and statistical weights $w_{N_e,j} = \exp[-\beta(E_{N_e,j}-\mu N_e)]/\sum_{N_e,j} \exp[-\beta(E_{N_e,j}-\mu N_e)]$. The thermal equilibrium of a grand canonical ensemble is then defined as that statistical density operator which minimizes the grand potential. Further details on the grand canonical ensemble formulation of a thermal electronic system can be found in Refs.~\cite{PPF+2011:exact,BCG2015:reduceddensitymatrixfunctional}. 

In practice, the electron-electron interaction complicates finding solutions to $\Psi_{N_e,j}$ and, hence, evaluating the grand potential as defined in Eq.~(\ref{eq.def.gp}). Instead, a solution is found in a computationally feasible manner within DFT~\cite{KS1965:selfconsistent} at finite electronic temperature~\cite{Mer1965:thermal,PPF+2011:exact,PPGB2014:thermal}. Here, a formally exact representation of the grand potential is given by
\begin{align}
	\label{eq:gp.ks}
	\begin{split}
		\Omega[n] &= T\s[n] - S\s[n]/\beta + U[n] + E\xc[n] \\
		&+ \int d\br\  n(\br;\ubR)\left[ v^{ei}(\br;\ubR) -\mu \right] + V^{ii}(\ubR)\,,
	\end{split}
\end{align}
where $T\s$ denotes the Kohn-Sham kinetic energy, $S\s$ the Kohn-Sham entropy, $U[n]$ the classical electrostatic interaction energy, $E\xc$ the exchange-correlation free energy which is the sum of the temperature-dependent exchange-correlation energy and the interacting entropy, $v^{ei}(\br;\ubR) = -\sum_\alpha Z_\alpha/|\br-\bR_\alpha|$ the electron-ion interaction, and $V^{ii}$ the ion-ion interaction.
The grand potential in Eq.~(\ref{eq:gp.ks}) is evaluated in terms of the electronic density defined by
\begin{equation}
	\label{eq:density.dft}
	n(\br; \ubR) = \sum_j f^\beta(\epsilon_j)\, |\phi_j(\br; \ubR)|^2\,,
\end{equation}
where the sum runs over the Kohn-Sham orbitals $\phi_j$ and eigenvalues $\epsilon_j$ that are obtained from the solving the Kohn-Sham equations
\begin{equation}
	\label{eq:ks.dft}
	\left[-\frac{1}{2}\nabla^2 + v\s(\br; \ubR)\right] \phi_j(\br; \ubR) = \epsilon_j \phi_j(\br; \ubR)\,,
\end{equation}
where $f^\beta(\epsilon_j) = (1+\exp[\beta(\epsilon_j-\mu)])^{-1}$ denotes the Fermi-Dirac distribution.
The Kohn-Sham framework is constructed such that a noninteracting system of fermions yields the same electronic density as the interacting many-body system of electrons defined by $\hat{H}^{BO}$. This is achieved by the Kohn-Sham potential $v\s(\br;\ubR) = \delta U[n]/\delta n(\br;\ubR) + \delta E\xc[n]/\delta n(\br;\ubR) + v^{ei}(\br;\ubR)$ which includes all electron-electron interactions on a mean-field level. While formally exact, approximations to $E\xc$ are applied in practice either at the electronic ground state~\cite{PGB2015:dft} or at finite temperature~\cite{BDHC2013:exchangecorrelation, KSDT2014:accurate, GDSM2017:initio}. 

The key quantity connecting the electronic and ionic degrees of freedoms is the total energy
\begin{align}
	\label{eq:fes}
	A^{BO}[n](\ubR) = \Omega_0[n] + \mu N_e\ .
\end{align}
This expression yields the Born-Oppenheimer potential energy surface at finite electronic temperature, when it is evaluated as a function of $\ubR$.
It provides the forces $\bF_\alpha = -\partial A^{BO}(\ubR)/\partial \bR_\alpha$ on the ions which are obtained from solving for the instantaneous, grand potential in thermal equilibrium $\Omega_0[n]$ for a given atomic configuration $\ubR$ via Eq.~(\ref{eq:ks.dft}). In doing so, Eq.~(\ref{eq:fes}) enables us to time-propagate the dynamics of the atomic configuration $\ubR$, in its simplest realization, by solving a Newtonian equation of motion
\begin{align}
	\label{eq:eom.ions.newton}
	{M_\alpha}\frac{d^2\bR_\alpha}{dt^2} = \bF_\alpha\ .    
\end{align}
This can be further extended, for example, to take into account the energy transfer between the coupled system of electrons and ions by introducing thermostats~\cite{BP1992:adiabaticity,GMCP1990:initio}. 

Despite its powerful utility for predicting thermodynamic and material properties on the atomic scale~\cite{Bec2014:perspective, LBB+2016:reproducibility} and providing benchmark data for the parameterization of IAPs for use in MD simulations~\cite{JNG2014:general}, applying DFT becomes computationally infeasible in applications towards the mesoscopic scale due to the computational scaling of Eq.~(\ref{eq:ks.dft}). The computational cost for data generation explodes exponentially with the number of chemical elements, thermodynamic states, phases, and interfaces~\cite{WCWT2019:datadriven} needed for the construction of IAPs.

\subsection{\label{ssec:ml-fes} Machine-learning Model of the Total Energy}
We now reformulate DFT in terms of the LDOS which is granted by virtue of the first Hohenberg-Kohn theorem~\cite{HK1964:inhomogeneous}. We then replace the DFT evaluation of the LDOS with a ML model for the LDOS in order to obtain all of the quantities required to evaluate the total energy while avoiding the computationally expensive solution of Eq.~(\ref{eq:ks.dft}). The LDOS is defined as a sum over the Kohn-Sham orbitals 
\begin{equation}
	\label{eq:ldos}
	D(\epsilon,\br;\ubR) = \sum_j |\phi_{j} (\br; \ubR)|^2\, \delta(\epsilon-\epsilon_j)\,,
\end{equation}
where $\epsilon$ denotes the energy as a continuous variable, and $\delta(\epsilon)$ is the Dirac delta function.
This enables us to rewrite Eq.~(\ref{eq:fes}) as 
\begin{align}
	\label{eq:FES.LDOS}
	\begin{split}
		A^{BO}[D](\ubR) & = E_{b}[D] - S\s[D]/\beta - U[D] \\
		& + E\xc[D] - V\xc[D]    +V^{ii}(\ubR)\,,    
	\end{split}
\end{align}
where $V\xc[D] = \int d\br\ n[D](\br; \ubR)\ \delta E\xc[D]/\delta n[D](\br;\ubR)$ denotes the potential energy component of the XC energy~\cite{DG1990:density}.  

The central advantage of this reformulation is that, by virtue of Eq.~(\ref{eq:FES.LDOS}), the total energy is expressed solely in terms of the LDOS. We emphasize this by explicitly denoting the functional dependence of each term on the LDOS.

Now let us turn to the individual terms in this expression.
The ion-ion interaction $V^{ii}(\ubR)$ is given trivially by the atomic configuration. 
The Hartree energy $U[D]=U[n[D]]$, the XC energy $E\xc[D]=E\xc[n[D]]$, and the potential XC energy $V\xc[D]=V\xc[n[D]]$ are all determined implicitly by the LDOS via their explicit dependence on the electronic density   
\begin{equation}
	\label{eq:n.from.ldos}
	n[D](\br; \ubR) = \int d\epsilon\ f^\beta(\epsilon)\, D(\epsilon,r; \ubR)\,.
\end{equation} 
Likewise, the band energy
\begin{equation}
	\label{eq:Eb.LDOS}
	E_{b}[D] = \int d\epsilon\ f^\beta(\epsilon)\, \epsilon\, \bar{D}[D](\epsilon;\ubR)
\end{equation}
and the Kohn-Sham entropy
\begin{align}
	\label{eq:entropy.dos}
	\begin{split}
		S\s[D] & = -\int d\epsilon\ 
		\left\{ f^\beta(\epsilon) \log[f^\beta(\epsilon)]\right.\\ 
		&\left. + [1-f^\beta(\epsilon)] \log[1-f^\beta(\epsilon)] \right\} \bar{D}[D](\epsilon;\ubR)
	\end{split}
\end{align}
are determined implicitly by the LDOS via their explicit dependence on the DOS 
\begin{equation}
	\label{eq:dos.from.ldos}
	\bar{D}[D](\epsilon;\ubR) = \int d\br D(\epsilon,\br;\ubR)\ .
\end{equation}

Eq.~(\ref{eq:FES.LDOS}) enables us to evaluate the total energy in terms of a single ML model that is trained on the LDOS and can predict the LDOS for snapshots unseen in training. This is advantageous, as it has been shown that learning energies from electronic properties, such as the density, is significantly more accurate~\cite{BVL+2017:bypassing} than learning them directly from atomic descriptors~\cite{BP2007:generalized,STT2014:sparse,LSP+2016:understanding}. 
Furthermore, developing an ML model based on the LDOS instead of the density avoids complications of prior models. For example, accuracy was lost due to an additional ML model needed to map densities to kinetic energies or noise was introduced by taking gradients of the ML kinetic energy models~\cite{SRH+2012:finding,SRH+2013:orbitalfree,SRMB2015:nonlinear,LKD2015:molecular}.
The two formulations in Eq.~(\ref{eq:fes}) and Eq.~(\ref{eq:FES.LDOS}) are mathematically equivalent. However, Eq.~(\ref{eq:FES.LDOS}) is more convenient in our ML-DFT workflow where we use the LDOS as the central quantity. To our knowledge, this work is the very first implementation of DFT at finite electronic temperature in terms of an ML model.

\subsection{\label{ssec:dft-data} Data Generation}

Our initial efforts have focused on developing an ML model for aluminum at ambient density (2.699 g/cc) and temperatures up to the melting point (933 K).  The training data for this model was generated by calculating the LDOS for a set of atomic configurations using the Quantum Espresso electronic structure code~\cite{GBB+2009:quantum, GAB+2017:advanced, GBB+2020:quantum}. The structures were generated from snapshots of equilibrated DFT-MD trajectories for 256 atom supercells of aluminum.  We calculated LDOS training data for 10 snapshots at room temperature, 10 snapshots of the crystalline phase at the melting point, and 10 snapshots of the liquid phase at the melting point.

Our DFT calculations used a scalar-relativistic, optimized norm-conserving Vanderbilt pseudopotential (Al.sr-pbesol.upf)~\cite{H2013:optimized} and the PBEsol exchange-correlation functional~\cite{PRC+2008:restoring}.  We used a 100 Rydberg plane-wave cutoff to represent the Kohn-Sham orbitals and a 400 Rydberg cutoff to represent densities and potentials.  These cutoffs resulted in a $200 \times 200 \times 200$ real-space grid for the densities and potentials, and we used this same grid to represent the LDOS.  The electronic occupations were generated using Fermi-Dirac smearing with the electronic temperature corresponding to the ionic temperature.  Monkhorst-Pack~\cite{MP1976:special} k-point sampling grids, shifted to include the gamma point when necessary, were used to sample the band structure over the Brillouin zone. The DFT-MD calculations that generated the snapshots used $2 \times 2 \times 2$ k-point sampling for the crystalline phase and $1 \times 1 \times 1$ k-point sampling for the liquid phase.

In contrast, the generation of the LDOS used $8 \times 8 \times 8$ k-point sampling.  The reason for this unusually high k-point sampling (for a 256 atom supercell) was to help overcome a challenge that arises when using the LDOS as the fundamental parameter in a ML surrogate for DFT.  For a periodic supercell, the LDOS is given as
\begin{equation}
	\label{eq:periodic_ldos}
	D(\epsilon,\br;\ubR) = \Omega_{BZ}^{-1} \int_{BZ} d\bk \; \sum_j^{N_{s}} |\phi_{j\bk} (\br; \ubR)|^2\, \delta(\epsilon-\epsilon_{j\bk})\,, 
\end{equation}
where the index $j$ now labels the Kohn-Sham orbitals at a particular $\bk$ value, and the vector index $\bk$ categorizes the Kohn-Sham orbitals by their transformation properties under the group of translation symmetries of the periodic supercell.  In particular,
\begin{equation}
	\phi_{j\bk} (\br; \ubR) = u_{j\bk} (\br; \ubR) \exp(i \bk \cdot \br),
\end{equation}
where $u_{j\bk} (\br; \ubR)$ is some function with the same periodicity as the periodically repeated atomic positions $\ubR$.

In Eq.~(\ref{eq:periodic_ldos}), the integral with respect to $\bk$ is taken over the first Brillouin zone (BZ), which has volume $\Omega_{BZ}$.  If this integral could be evaluated exactly, the resulting LDOS would be a continuous function of $\epsilon$ with Van Hove singularities~\cite{LVH1953:VanHove} of the form $\sqrt{\epsilon - \epsilon'}$ or $\sqrt{\epsilon' - \epsilon}$ at energies $\epsilon'$ that correspond to critical points of the band structure $\epsilon_{j\bk}$. However, when this integral is represented as a summation over a finite number of k-points ${\bk_1, \ldots, \bk_{N_k}}$, as in the Monkhorst-Pack~\cite{MP1976:special} approach used in our calculations, it is necessary to replace the Dirac delta function $\delta(\epsilon-\epsilon_{j\bk})$ appearing in Eq.~(\ref{eq:periodic_ldos}) with a finite-width representation $\tilde{\delta}(\epsilon-\epsilon_{j\bk})$.  The LDOS used to train our ML model is then obtained by evaluating
\begin{equation}
	\label{eq:discrete_ldos}
	D(\epsilon,\br;\ubR) = N_k^{-1} \sum_k^{N_k} \sum_j^{N_{s}} \; |\phi_{j\bk_k} (\br; \ubR)|^2\, \tilde{\delta}(\epsilon-\epsilon_{j\bk_k}) 
\end{equation}
on a finite grid of $\epsilon$ values.  In our calculations, we use an evenly spaced $\epsilon$ grid with a spacing of 0.1 eV, a minimum value of -10 eV, and a maximum value of +15 eV. Hence, the data for each grid point consists of a vector of 250 scalar values. Kohn-Sham orbitals with eigenvalues significantly outside the energy range where we evaluate the LDOS make negligible contributions to the calculated LDOS values, and ideally we would include a sufficient number $N_s$ of the lowest Kohn-Sham orbitals in Eq.~(\ref{eq:periodic_ldos}) and Eq.~(\ref{eq:discrete_ldos}) to span this energy range.  However, due to memory constraints on our $8 \times 8 \times 8$ k-point DFT calculations, we were only able to calculate $N_s = 576$ orbitals. At ambient density, this provides an accurate LDOS up to $\epsilon \approx 10$ eV, and the relevant Fermi levels are at $\mu \approx 7.7$ eV.  The tails of $f^\beta(\epsilon)$ should be negligible above $\epsilon \approx 10$ eV for $T \lessapprox 2500$ K. At higher electronic temperatures, it would become necessary to increase $N_S$ when generating the training data in order to get an accurate LDOS over a wider energy range.

\begin{figure}[ht!]
	\centering
	\includegraphics[scale=0.27]{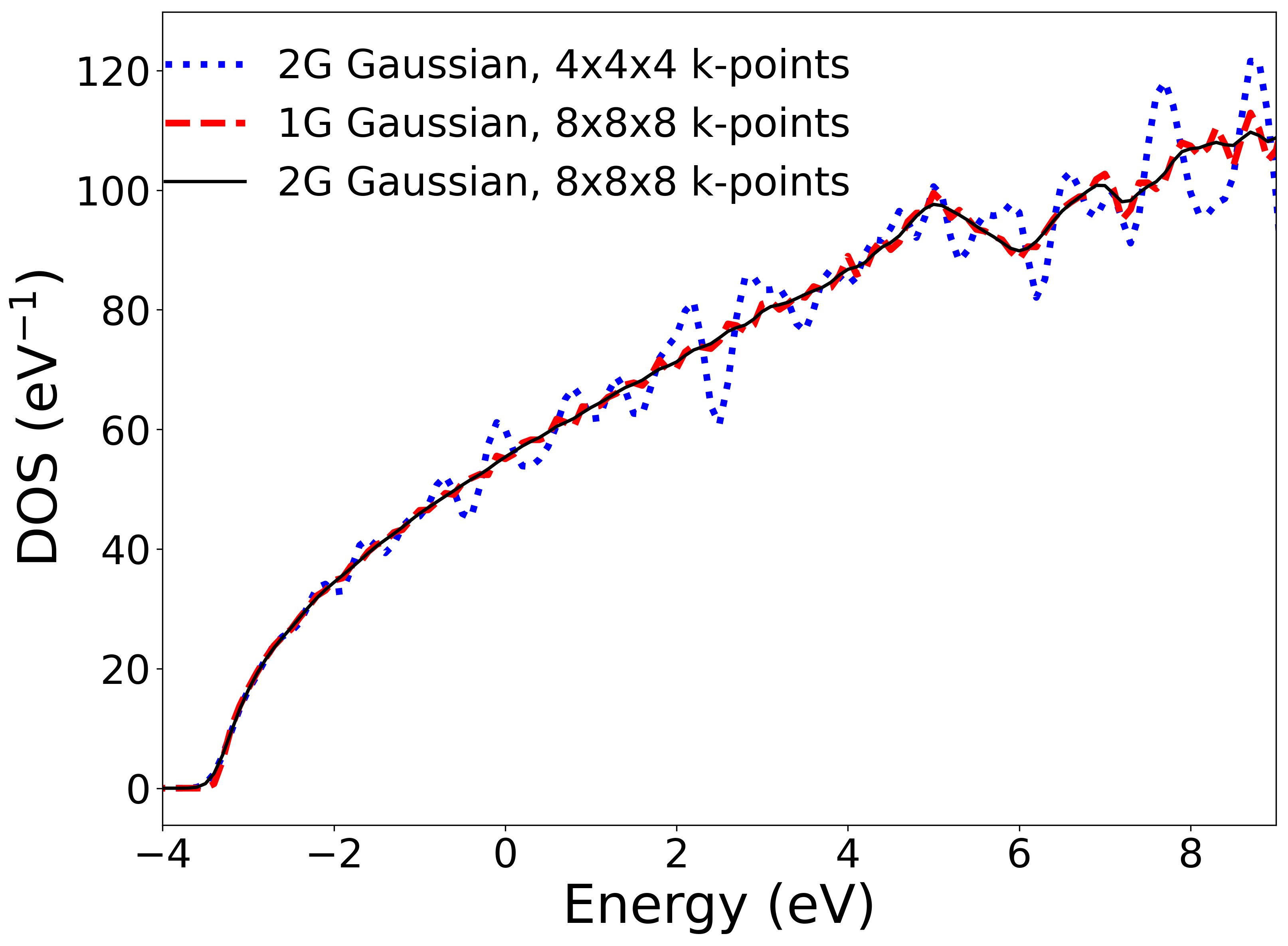}
	\includegraphics[scale=0.27]{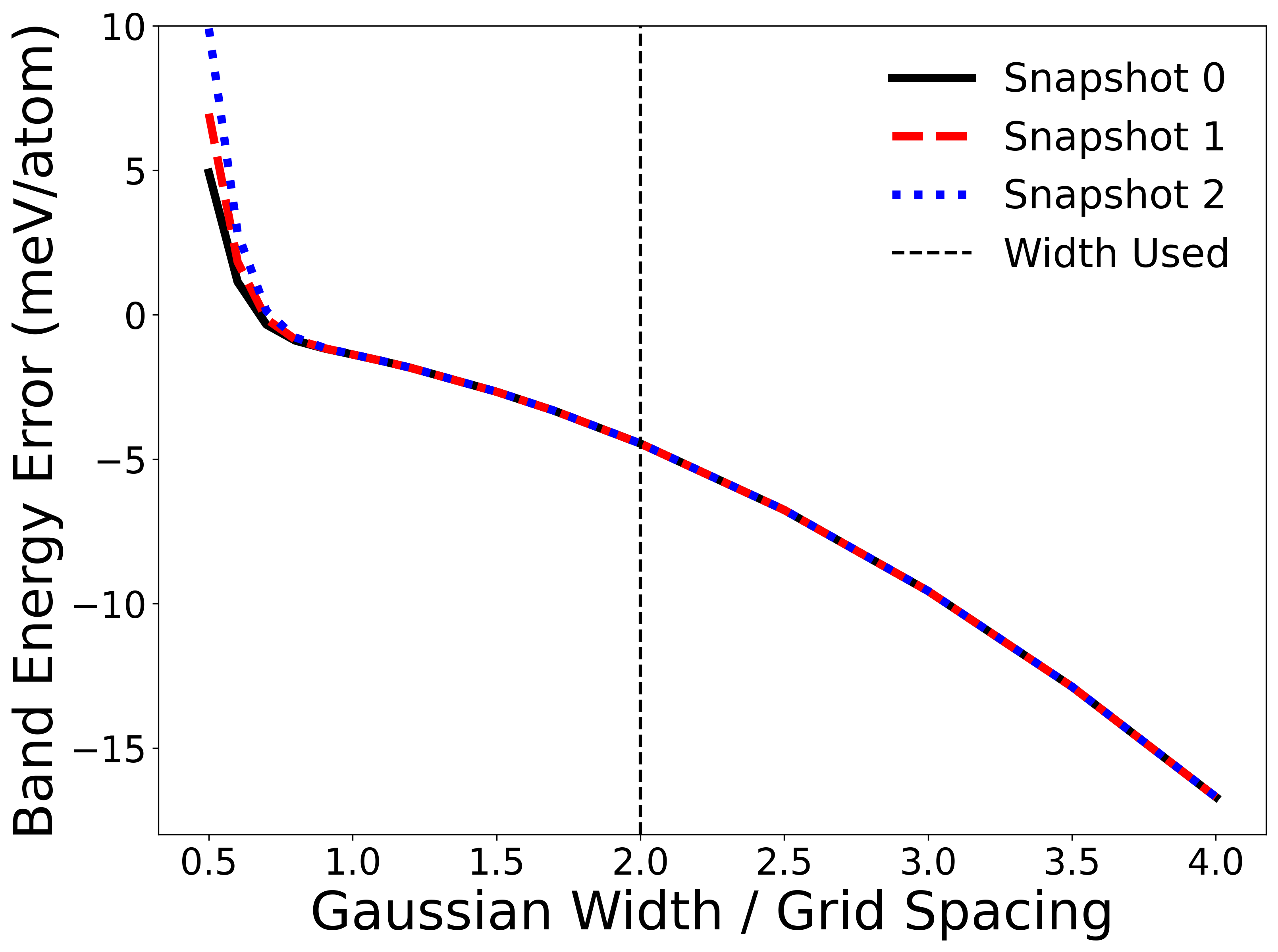}
	\caption{The DOS calculated for one of the 256-atom aluminum snapshots at ambient temperature (298~K) using different Gaussian smearing widths and k-point sampling schemes (above) and the error in the band energy as a function of the Gaussian smearing width for three different snapshots (below).  In the upper panel, ``1G Gaussian" indicates results obtained from a Gaussian $\tilde{\delta}(x)$ with $\sigma = 0.1$ eV (equal to the $\epsilon$ grid spacing), while ``2G Gaussian" indicates that $\sigma = 0.2$ eV (twice the $\epsilon$ grid spacing) was used.  Results are shown for $4 \times 4 \times 4$ and $8 \times 8 \times 8$ Monkhorst-Pack k-point sampling in the upper panel, and $8 \times 8 \times 8$ sampling in the lower panel.}
	\label{fig:ldos_definitions}
\end{figure}

The challenge mentioned above arises in the choice of the delta function representation $\tilde{\delta}$ and the k-point sampling scheme ${\bk_1, \ldots, \bk_{N_k}}$ for a given $\epsilon$ grid.  If the function $\tilde{\delta}$ is too wide, it leads to systematic errors in the LDOS and derived quantities such as the density, band energy and total energy. However, if $\tilde{\delta}$ is too narrow or too few k-points are used to sample the BZ, then only a few values of $\epsilon_{j\bk_k}$ make substantial contributions to the LDOS at each point on the $\epsilon$ grid, and there is a large amount of noise in the resulting LDOS. 

Figure~\ref{fig:ldos_definitions} demonstrates these trade-offs for a Gaussian representation $\tilde{\delta}(x) = \exp{\left(-x^2/\sigma^2\right)}/\sqrt{\pi\sigma^2}$ by plotting the DOS $\bar{D}[D](\epsilon;\ubR)$ in the upper panel and the error in the band energy calculated from the DOS in the lower panel.  The top panel shows that using fewer k-points or a narrower $\tilde{\delta}$ leads to unphysical noise in the calculated DOS, while the bottom panel shows that using a wider $\tilde{\delta}$ leads to an increasing error in the band energy.  The lower panel also shows that the noise in the DOS that arises with a narrower $\tilde{\delta}$ is reflected in an increasing variability in the band energy error.

Since a larger k-point sampling allows a narrower $\tilde{\delta}$ to be used without introducing excessive noise into the LDOS, we can minimize the errors in quantities calculated from the LDOS by using as many k-points as possible.  In the LDOS calculations used to train our ML model, we used $8 \times 8 \times 8$ k-point sampling, which was the largest k-point sampling that was computationally tractable for our 256 atom cells.  We then chose a Gaussian $\tilde{\delta}$ with $\sigma = 0.2$ eV (indicated in the lower panel of Figure~\ref{fig:ldos_definitions}), which gives a relatively small and very consistent error in derived quantities. For example, the average error in the band energies calculated from the resulting LDOS (relative to fully converged DFT) is -4.483 meV/atom when the average is taken over all of our snapshots.  The average error becomes -4.466, -4.500, and -4.483 meV/atom when the average is restricted to the 298K, 933K solid, and 933K liquid snapshots, respectively.  These results show that the error changes very little between different temperatures and structures.  These errors are also very consistent for different structures at the same temperature and phase.  In particular, the root mean square variations in the errors are 0.004, 0.011, and 0.017 meV/atom for the 298K, 933K solid, and 933K liquid snapshots respectively. Our choices for the $\epsilon$ grid spacing and the width of the Gaussian broadening turn out to be similar to those made in Ref.~\cite{mahmoud2020learning}.

In electronic structure calculations, consistent errors in calculated energies (for example, the errors due to incomplete convergence with respect to the plane-wave cutoffs in traditional DFT calculations) are generally not a problem.  Instead, quantities of interest almost always involve the difference between two energies, and thus it is the variation of the errors between different structures that affects actual results. With the k-point sampling and the choice of $\tilde{\delta}$ described above, the errors differ by only about one-thousandth of the room-temperature thermal energy, and such consistent errors are very unlikely to impact quantities of interest.  Furthermore, these errors are independent of the errors introduced by using ML to approximate the LDOS. In particular, a perfect ML approximation would reproduce the results calculated from the DFT LDOS (defined in the same manner as in the generation of our training data) rather than the results calculated by exact DFT. Finally, we believe that further work will be able to identify methods that use DFT to calculate even more accurate LDOS training data with less computational effort than our $8 \times 8 \times 8$ k-point calculations. Some ideas include Fourier interpolation of the Kohn-Sham bands and orbitals, or an approach analogous to the tetrahedron method~\cite{BJA1994:improved}. The remainder of this paper will focus on the effects of approximating the LDOS with ML. In light of the above considerations, errors will be quoted relative to results calculated from the DFT LDOS rather than exact DFT.

\subsection{\label{ssec:eval-data} Evaluation of Energies from the Local Density of States}
A key step in calculating the total energy defined in Eq.~(\ref{eq:FES.LDOS}) from the LDOS is the evaluation of several integrals in Eqs.~(\ref{eq:n.from.ldos}), (\ref{eq:Eb.LDOS}), and (\ref{eq:entropy.dos}) that have the form
\begin{equation}
	\label{eq:form_of_integrals}
	I = \int d\epsilon\ g(\epsilon)\, D(\epsilon)\ ,
\end{equation} 

for some function $g(\epsilon)$. The evaluation of Eq.~(\ref{eq:form_of_integrals}) poses a challenge, because $g(\epsilon)$ changes rapidly compared to $D(\epsilon)$ in several cases of interest. We provide a solution in terms of an analytical integration, as illustrated in Appendix~\ref{app:ldos-eval}. Given our analytical integration technique, it is straightforward to evaluate the total energy using Eq.~(\ref{eq:FES.LDOS}) and the standard methods of DFT in an accurate and numerically stable manner.

\subsection{\label{ssec:fingerprint} Fingerprint Generation}
We assume that the LDOS at any point in space can be approximated by a function that depends only on the positions and chemical identities of atoms within some finite neighborhood of the point. In order to approximate this function using ML, we need to construct a fingerprint that maps the neighborhood of any point to a set of scalar values called \emph{descriptors}. On physical grounds, good descriptors must satisfy certain minimum requirements: (i) invariance under permutation, translation, and rotation of the atoms in the neighborhood and (ii) continuous differentiable mapping from atomic positions to descriptors, especially at the boundary of the neighborhood.  While these requirements exclude some otherwise appealing choices, such as the Coulomb matrix~\cite{Rupp2012}, the space of physically valid descriptors is still vast. In the context of ML-IAPs, the construction of good atomic neighborhood descriptors has been the subject of intense research. Recent work by Drautz~\cite{drautz2019} has shown that many prominent descriptors~\cite{behler2007, bartok2010, thompson2015} belong to a larger family that are obtained from successively higher order terms in an expansion of the local atomic density in cluster integrals.  The SNAP bispectrum descriptors~\cite{TST+2015:spectral} that we use in this work correspond to clusters of three neighbor atoms yielding four-body descriptors.

In contrast to prior work on ML-IAPs in which descriptors are evaluated on atom-centered neighborhoods, here we evaluate the SNAP descriptors on the same $200 \times 200 \times 200$ Cartesian grid points at which we evaluated the LDOS training data (see Eq.~(\ref{eq:discrete_ldos}) above). For
each grid point, we position it at the origin and define local atom positions in the neighborhood
relative to it. The total density of neighbor atoms is represented as a sum of $\delta$-functions in a three-dimensional space:
\begin{equation}
	\rho ({\bf r}) = \delta({\bf 0}) + \!\!\!\!\!\!  \sum_{r_{k} < R_{cut}^{\nu_k}} \!\!\!\!\!\!  {f_c(r_{k}; R_{cut}^{\nu_k}) w_{\nu_k} \delta({\bf r}_{k})}\,,
	\label{eq:totaldensity}
\end{equation}
where ${\bf r}_{k}$ is the position of the neighbor atom $k$ of element $\nu_k$ relative to the grid point.  The $w_{\nu}$ coefficients are dimensionless weights that are chosen to distinguish atoms of different chemical elements $\nu$, while a weight of unity is assigned to the location of the LDOS point at the origin.  The sum is over all atoms $k$ within some cutoff distance $R_{cut}^{\nu_k}$ that depends on the chemical identity of the neighbor atom.  The switching function $f_c(r; R_{cut}^{\nu_k})$ ensures that the contribution of each neighbor atom goes smoothly to zero at $R_{cut}^{\nu_k}$.  Following Bart{\'{o}}k \emph{et~al.}~\cite{bartok2010}, the radial distance $r_{k}$ is mapped to a third polar angle $\theta_0$ defined by
\begin{equation}
	\theta_0 = \theta_0^{max}\frac{r_{k}}{R_{cut}^{\nu_k}}\ .
\end{equation}
The additional angle $\theta_0$ allows the set of points ${\bf r}_{k}$ in the 3D ball of possible neighbor positions to be mapped on to the set of points $(\theta, \phi, \theta_0)$ on the unit 3-sphere.  
The neighbor density function can be expanded in the basis of 4D hyperspherical harmonic functions $\bU^j$
\begin{equation}
	\rho({\bf r}) = \sum_{j=0,\frac{1}{2},\ldots}^{\infty}\bu_j\cdot\bU_j(\theta_0,\theta,\phi)\,,
\end{equation}
where $\bu_j$ and $\bU_j$ are rank $(2j+1)$ complex square matrices, $\bu_j$ are Fourier expansion coefficients given by the inner product of the neighbor density with the basis functions $\bU_j$ of degree $j$, and the $\cdot$ symbol indicates the scalar product of the two matrices.  Because the neighbor density is a weighted sum of $\delta$-functions, each expansion coefficient can be written as a sum over discrete values of the corresponding basis function
\begin{eqnarray}
	\bu_j &=& \bU_j({\bf 0}) + 
	\!\!\!\!\!\!
	\sum_{r_{k} < R_{cut}^{\nu_k}} {f_c(r_{k}; R_{cut}^{\nu_k}) w_{\nu_k} \bU_j(\theta_0,\theta,\phi)}\ .
\end{eqnarray}
The expansion coefficients $\bu_j$ are complex-valued and they are not directly useful as descriptors because they are not invariant under rotation of the polar coordinate frame.  However, scalar triple products of the expansion coefficients 
\begin{equation}
	B_{j_1j_2j}  =  \bu_{j_1}\otimes_{j_1j_2j}\bu_{j_2}\cdot(\bu_j)^*
\end{equation}
are real-valued and invariant under rotation~\cite{bartok2010}.
The symbol $\otimes_{j_1j_2j}$ indicates a Clebsch-Gordan product of matrices of rank $j_1$ and $j_2$ that produces a matrix of rank $j$, as defined in our original formulation of SNAP~\cite{thompson2015}.  
These invariants are the components of the bispectrum.  They characterize the strength of density correlations at three points on the 3-sphere.  The lowest-order components describe the coarsest features of the density function, while higher-order components reflect finer detail.  The bispectrum components defined here have been shown to be closely related to the 4-body basis functions of the Atomic Cluster Expansion introduced by Drautz~\cite{drautz2019}.

For computational convenience, the LAMMPS implementation of the SNAP descriptors on the grid includes all unique bispectrum components $B_{j_1j_2j}$ with indices no greater than $J_{max}$. For the current study we chose $J_{max}=5$, yielding a fingerprint vector consisting of 91 scalar values for each grid point.  The cutoff distance and element weight for the aluminum atoms were set to 4.676~\AA~and $1.0$, respectively.

\subsection{\label{ssec:ml_workflow} Machine-learning Workflow and Neural Network Model}

\begin{table}[!t]
	\small
	\begin{center}
		\caption{ML-DFT model training workflow.}
		\begin{ruledtabular}
			\begin{tabular}{ p{0.03\columnwidth} p{0.9\columnwidth} }
				{\bf 1.} & {\bf Training data generation} \\
				(i.)   & Run MD simulation to obtain atomic configurations at a specific temperature and density. \\
				(ii.)   & Run Quantum Espresso DFT calculation on the atomic configurations. \\
				(iii.)   & Perform Quantum Espresso post-processing of the Kohn-Sham orbitals to obtain the LDOS on the Cartesian grid. \\
				(iv.)   & Generate SNAP fingerprints on a uniform Cartesian grid from the atomic configurations. \\
				\midrule
				{\bf 2.} & {\bf ML-DFT model training} \\
				(i.)   & Training/validation set, ML model, and initial hyperparameter selection. \\
				(ii.)   & Normalization/standardization of the training data. \\
				(iii.)   & Train ML-DFT model to convergence. \\
				(iv.)   & Adjust hyperparameters for increased accuracy and repeat training for a new model.\\
				\midrule
				{\bf 3.} & {\bf ML-DFT analysis} \\
				(i.)   & Evaluate unseen test snapshots on trained and tuned ML-DFT models. \\
				(ii.)   & Calculate quantities of interest from the LDOS predicted by ML-DFT.\\
			\end{tabular}
		\end{ruledtabular}
		
		\label{tab:workflow}
	\end{center}
\end{table}

Neural network models, especially deep neural networks, are profoundly successful in numerous tasks, such as parameter estimation~\cite{schmidt2019recent, cosmoflow2018}, image classification~\cite{image2016, Ziletti2018InsightfulCO}, and natural language processing~\cite{Young2018RecentTI, bert2019}. Moreover, neural networks may act as function approximators~\cite{approx1993, deeplearning2016}. A typical \emph{feed-forward} neural network is constructed as a sequence of transformations, or \emph{layers}, with the form: 
\begin{equation}
	v^{\ell + 1}_{(n,1)} = \varphi({\bf W}^{\ell}_{(n,m)} v^{\ell}_{(m,1)} + b^{\ell}_{(n,1)}),
\end{equation}
where $v^{\ell}_{(m,1)}$ is the intermediate column-vector of length $m$ at layer $\ell$, ${\bf W}^{\ell}_{(n,m)}$ is a matrix of size $n \times m$, $b^{\ell}_{(n,1)}$ is a bias vector of length $n$, and $v^{\ell + 1}_{(n, 1)}$ is the next column-vector of length $n$ in the sequence. The first and the last layers accept \emph{input} vectors, $v^{0}$, and produce \emph{output} predictions, $v^{L}$, respectively. There can be any arbitrary number $L-1$ of \emph{hidden} layers constructed between the input layer and the output layer. After each layer, a non-linear {\it activation} unit $\varphi$ transforms its input. The composition of these layers is expected to learn complex, non-linear mappings. 

Given a dataset of $N_s$ input vectors and target output vectors, ($v^{0}$, $\bar{v}^{L}$), we seek to find the optimal $W^{\ell}$ and $b^{\ell}$, $\forall \ell \in [0, L-1]$, such that the neural network minimizes the root mean squared loss between the neural network predictions $v^{L}$ and targets $\bar{v}^{L}$. In this work, the network predicts 
LDOS at each grid point using the SNAP fingerprint as input. The specific feed-forward networks we consider have input fingerprint vectors of length 91 scalar values and output LDOS vectors of length 250 scalar values. The grid points are defined on a uniform $200 \times 200 \times 200$ Cartesian grid, or 8 million total points per snapshot. 

Neural network models are created in two stages: {\it training} and {\it validation}, using separate training and validation datasets. 
The training phase performs gradient-based updates, or \emph{backpropagation}~\cite{deeplearning2016}, to $W^{\ell}$ and $b^{\ell}$ using a small subset, or {\it mini-batch}, of column vectors in the training dataset. Steps taken in the direction of the gradient are weighted by a user provided \emph{learning rate} (LR). Successive updates incrementally optimize the neural network. An {\it epoch} is one full pass through the training dataset for the training phase. After each epoch, we perform the validation phase, which is an evaluation of the current model's accuracy on a validation dataset. The training process continues alternating between training and validation phases until an {\it early stopping} termination criterion~\cite{Prechelt1996EarlySW} is satisfied. Early stopping terminates the learning process when the root mean squared loss on the validation dataset does not decrease for a successive number of epochs, or {\it patience} count. Increasing validation loss indicates that the model is starting to over-fit the training dataset and will have worse generalization performance if the training continues.

Next, the hyperparameters, which determine the neural network architecture and optimizer specifications, must be optimized to maximize model accuracy (and minimize loss). Due to the limited throughput, we make the assumption that all hyperparameters are independent. Independence allows the hyperparameter tuning to optimize one parameter at a time, greatly reducing the total rounds of training. We then ensure local optimality of the hyperparameter minimizer by direct search~\cite{lewis2000191}.

The ML-DFT model uses as input vectors the SNAP fingerprints defined at real-space grid points. The target output vectors of the ML-DFT model are the Quantum Espresso generated LDOS vectors defined at the corresponding real-space grid points. Importantly, each fingerprint and LDOS vector is treated as independent of any other grid point. Further, the training data may include multiple aluminum snapshots, in which case the number of grid points is equal to the number of snapshots times the number of grid points per snapshot. No distinction is made in the model by grid points belonging to separate snapshots. The validation snapshot is always a full set of grid points from a single snapshot. 

For inference, all remaining snapshots are used to test the generalization of the ML-DFT model. Consider a single snapshot where we wish to predict the total energy. At each grid point, a SNAP fingerprint vector is used to predict an LDOS vector. Using the steps in Section~\ref{ssec:eval-data} and Appendix~\ref{app:ldos-eval}, the full set of ML predicted LDOS vectors produce a total energy.

In summary, the ML-DFT training workflow is provided in Table~\ref{tab:workflow}. The ML-DFT models are built on top of the PyTorch~\cite{paszke2017automatic} framework and the Numpy~\cite{oliphant2006guide} and Horovod~\cite{sergeev2018horovod} packages. PyTorch provides the necessary infrastructure for neural network construction and efficiently running the training calculations on GPU machines. The Numpy package contains many performant tools for manipulating tensor data, and the Horovod package enables data-parallel training for MPI-based scalability. 

\section{\label{sec:results} Results}
\begin{figure*}[htpb]
	\centering
	\includegraphics[scale=0.49]{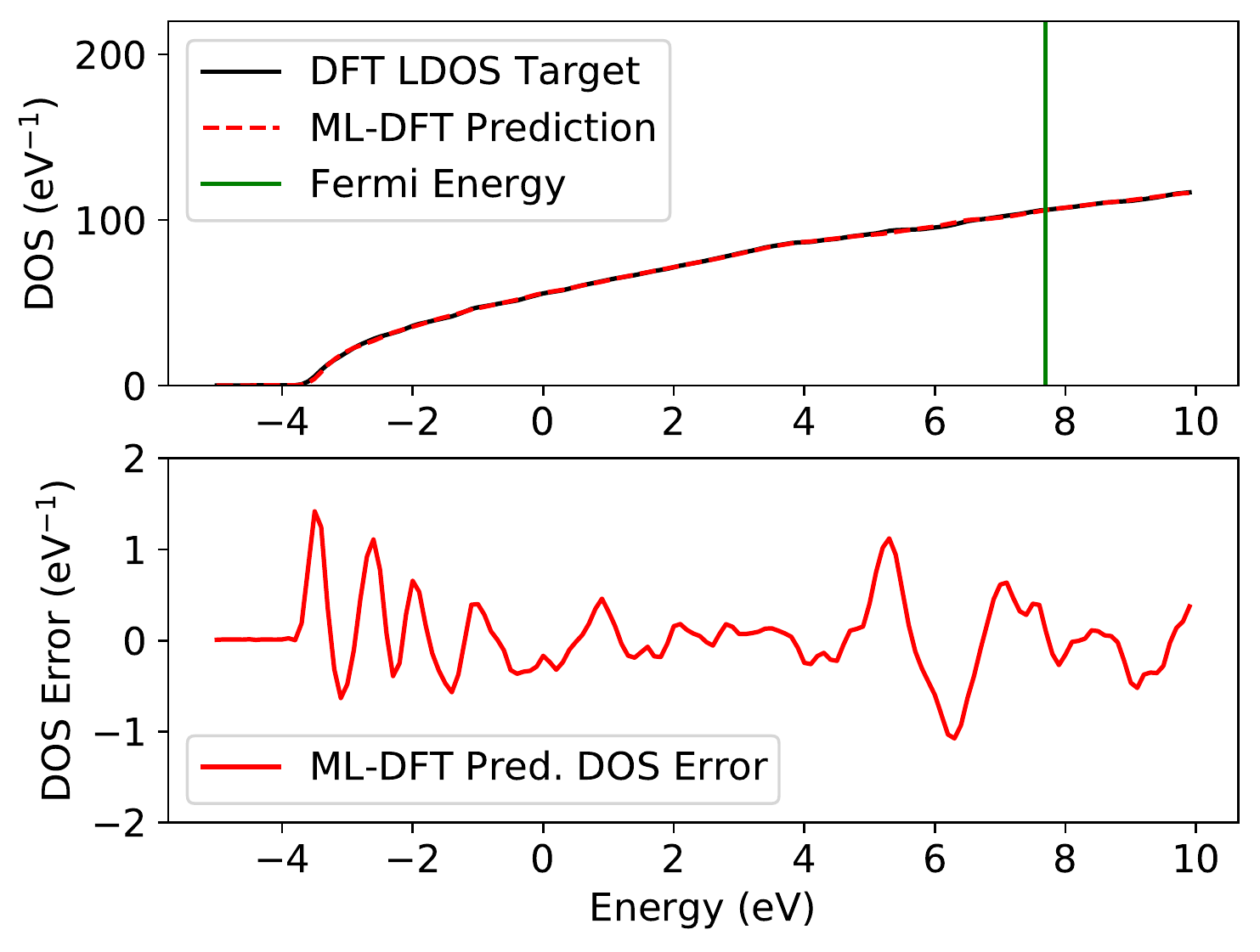}
	\includegraphics[scale=0.49]{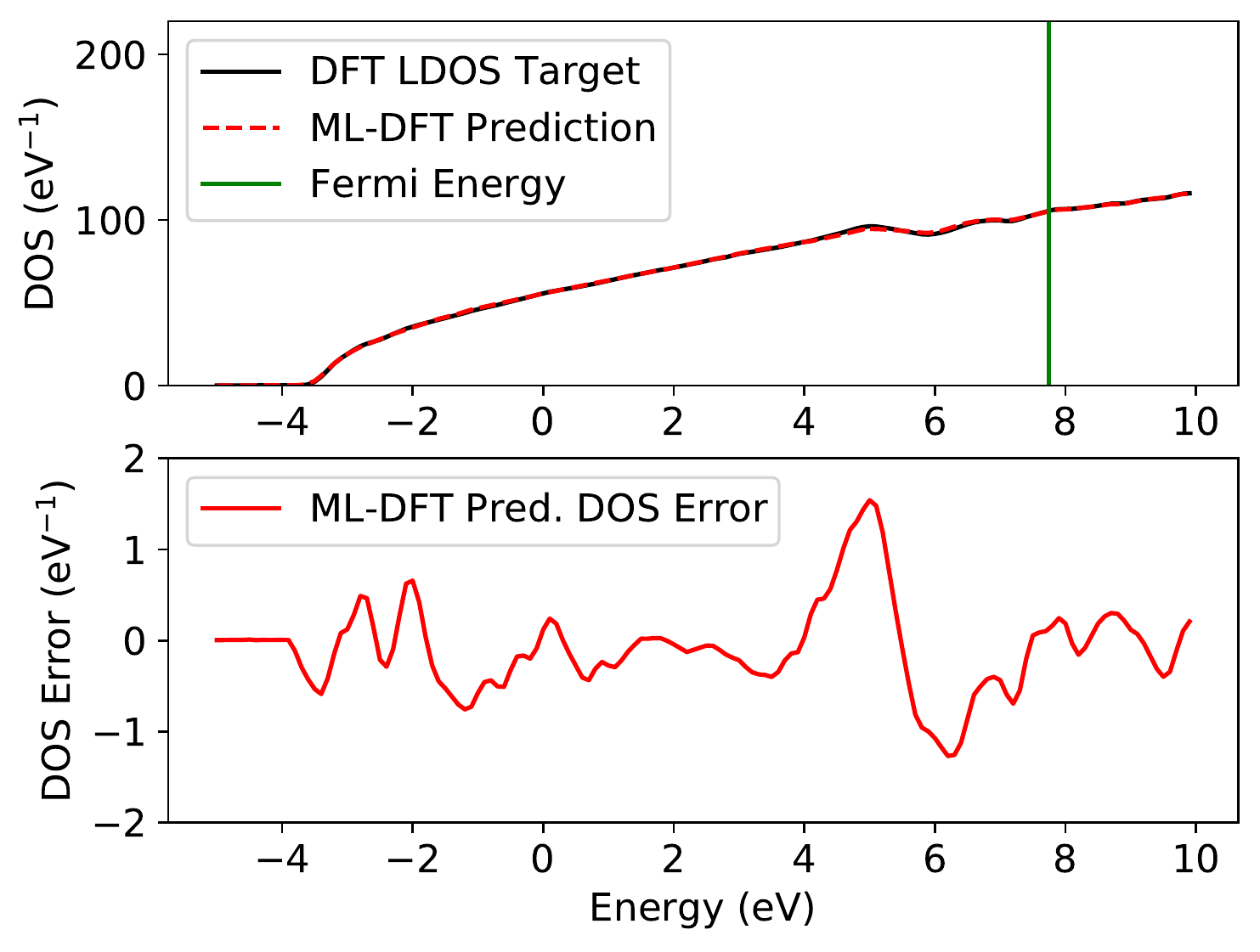}
	\caption{Liquid (left) and solid (right) test snapshot DOS comparisons for the ML-DFT model trained on 6 liquid and 6 solid snapshots at 933~K. The vertical lines indicate the Fermi energy which is 7.689~eV for the liquid snapshot and 7.750~eV for the solid snapshot.}
	\label{fig:hybrid_dos}
\end{figure*}

\begin{figure*}[htpb]
	\centering
	\includegraphics[scale=0.36]{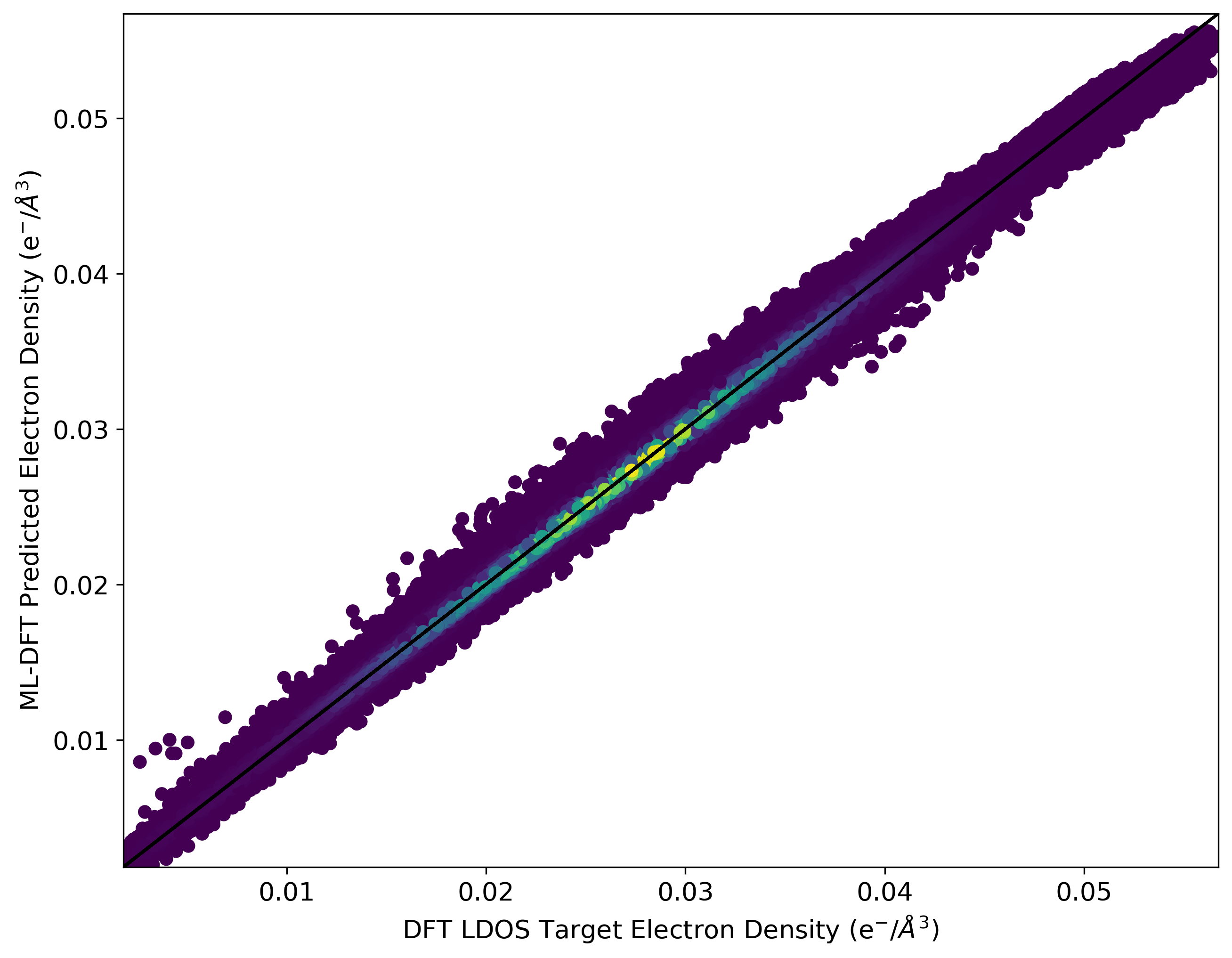}
	\includegraphics[scale=0.36]{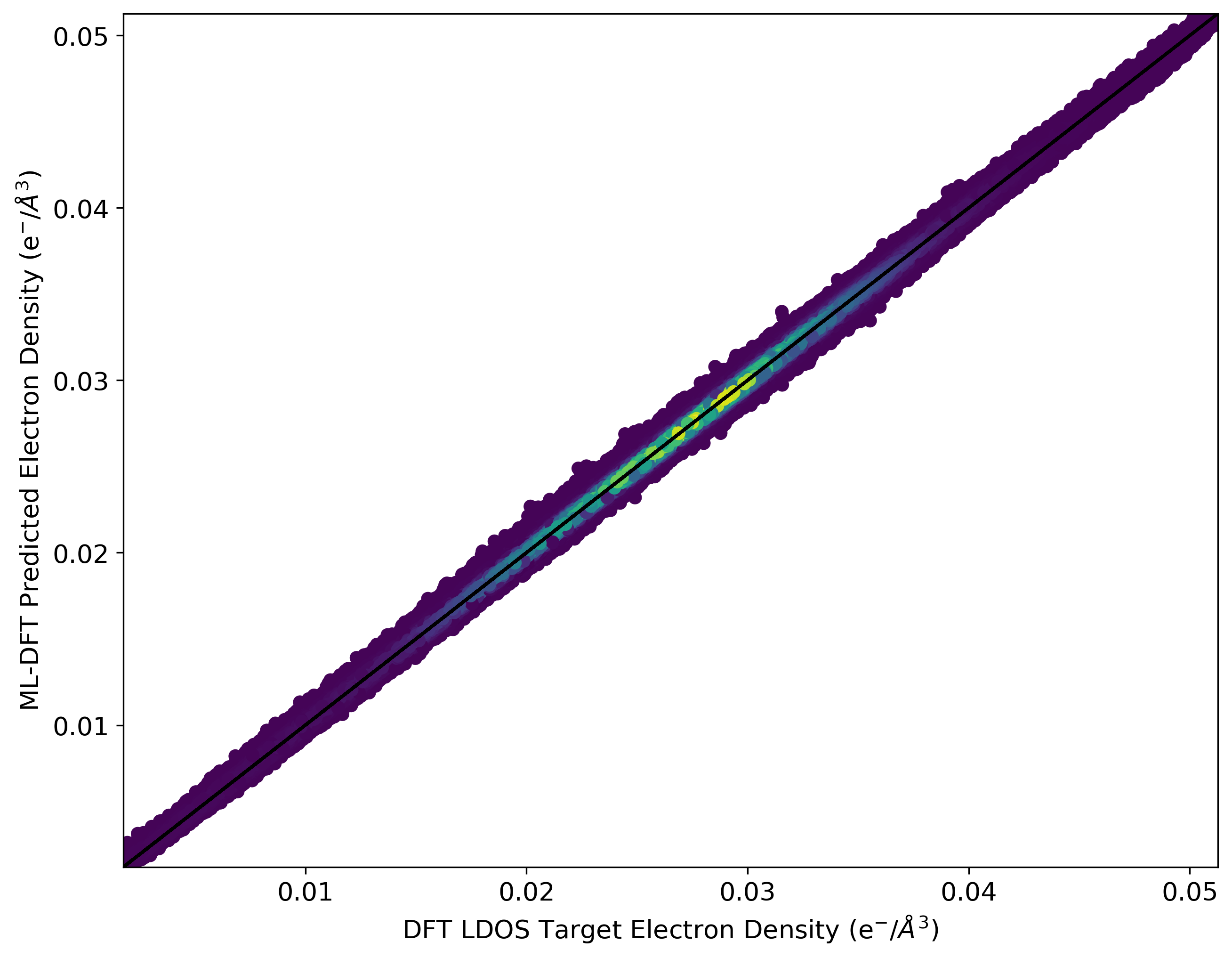}
	\includegraphics[scale=0.09]{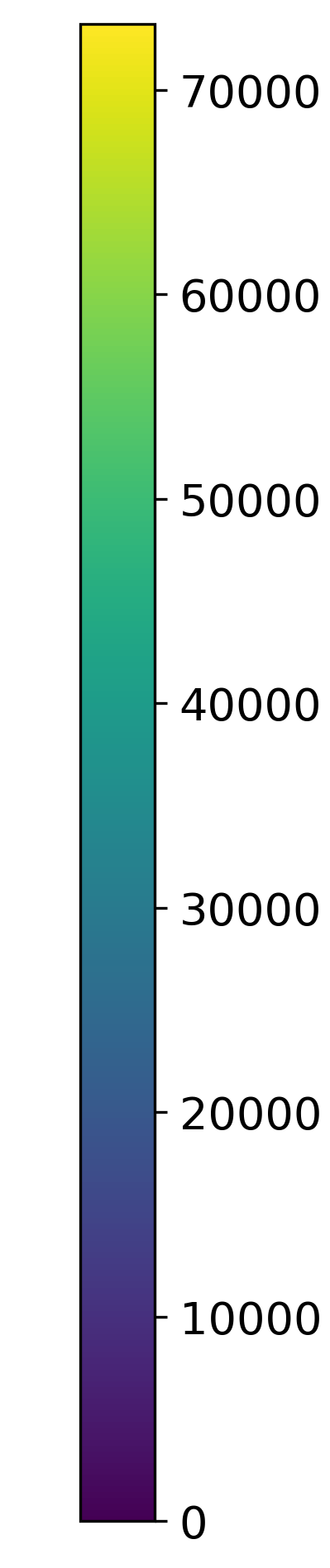}
	\caption{Liquid (left) and solid (right) test snapshot electron density comparisons for the ML-DFT model trained on 6 liquid and 6 solid snapshots at 933~K. The maximum absolute errors for the liquid and solid test snapshot are 0.00601~$e^-/\rm\AA^3$ and 0.00254~$e^-/\rm\AA^3$, respectively, mean absolute errors are 0.000284~$e^-/\rm\AA^3$ and 0.000167~$e^-/\rm\AA^3$, respectively and mean absolute percentage errors are 1.12\% and 0.66\%, respectively. Marker color indicates the frequency.}
	\label{fig:hybrid_density}
\end{figure*}

The difficulty of predicting the electronic structure of materials is strongly dependent on the diversity of local atomic configurations considered in training and testing. 
During the development stage of this work we focused on the relatively simple case of solid aluminum at ambient temperature and density (298~K, 2.699~g/cm$^3$). 
We then turned our attention to the more challenging case of solid and liquid aluminum near the melting point (933~K, 2.699~g/cm$^3$) and we present results for that case here. 

Specifically, we train and tune a collection of unique ML-DFT models for each temperature. In the 298~K case, we train a single model using 1 snapshot, validate on 1 snapshot, and test on 8 snapshots. At 933~K, we train 8 models with different combinations of liquid and solid snapshots as the training set. The first 3 cases use 4 training snapshots with either 4 liquid snapshots, 4 solid snapshots, or 2 liquid and 2 solid snapshots. The second 3 cases, similarly, use 8 training snapshots with either 8 liquid, 8 solid, or 4 liquid and 4 solid snapshots. Finally, there are two models that use 8 or 12 training snapshots and a validation set that include both liquid and solid snapshots. Each 933~K model created is then tested on the remaining 933~K snapshots. In order to study the variability in LDOS between the three groups (298~K solid, 933~K solid, and 933~K liquid), we reduce the dimensionality of the LDOS datasets and study them using principal component analysis (PCA).

Further technical details on fingerprint generation, neural network architecture, ML training, optimization of hyper-parameters, and the variability in LDOS outputs are given in Appendix~\ref{app:ml-model-details}.
We provide results for the ambient case in Appendix~\ref{app:Al-solid-298K}.

\subsection{\label{ssec:ml_pred_ldos} Local Density of States Predictions for Solid and Liquid Aluminum from ML-DFT}

First we assess the accuracy of ML-DFT for spatially resolved and energy resolved quantities. To this end, Figure~\ref{fig:hybrid_dos} illustrates the DOS and its errors in the relevant energy range from -5 to 10 eV, where the Fermi energy is at 7.689 eV in the liquid and 7.750 eV in the solid phase.  
The DOS prediction is computed from Eq.~(\ref{eq:dos.from.ldos}) using the ML-DFT predicted LDOS. In both liquid (left) and solid (right) phases, the ML-DFT model reproduces the DOS to very high accuracy. The illustration of the errors (bottom panels) confirms the ML-DFT model's accuracy and shows the absence of any unwanted error cancellation over the energy grid.  Figure~\ref{fig:hybrid_dos} also shows that the Van Hove singularities~\cite{LVH1953:VanHove} at and above $\approx 5$ eV, which show up strongly in the 298K DOS in Figure~\ref{fig:ldos_definitions}, are noticeably smeared out by thermal fluctuations in the atomic positions in the 933~K solid-phase DOS and essentially gone in the 933~K liquid. 
Figure~\ref{fig:hybrid_density} shows the DFT target electronic density versus the predicted electronic density for both liquid (left) and solid (right) phases at each point on the Cartesian grid. The frequency of the predicted results is indicated by the marker color. The ML-DFT prediction is computed from Eq.~(\ref{eq:n.from.ldos}) using the ML-DFT predicted LDOS. The alignments of the predictions along the diagonal exhibit small standard deviations for the liquid and solid test snapshot.
It illustrates how well the ML-DFT model reproduces the electronic density in a systematic manner with high accuracy, despite the fact that the electronic structure is qualitatively very distinct in the solid and liquid phases of aluminum. Furthermore, maximum absolute errors for the liquid and solid test snapshot are 0.00601~$e^-/\rm\AA^3$ and 0.00254~$e^-/\rm\AA^3$, respectively. The mean absolute errors are an order of magnitude lower at 0.000284~$e^-/\rm\AA^3$ and 0.000167~$e^-/\rm\AA^3$, respectively. The mean absolute percentage errors are 1.12\% and 0.66\%, respectively.

\subsection{\label{ssec:ml_pred_single-phase-model}Single-phase Solid and Liquid Aluminum ML-DFT Models}

\begin{figure}[htpb]
	\centering
	\includegraphics[width=\columnwidth]{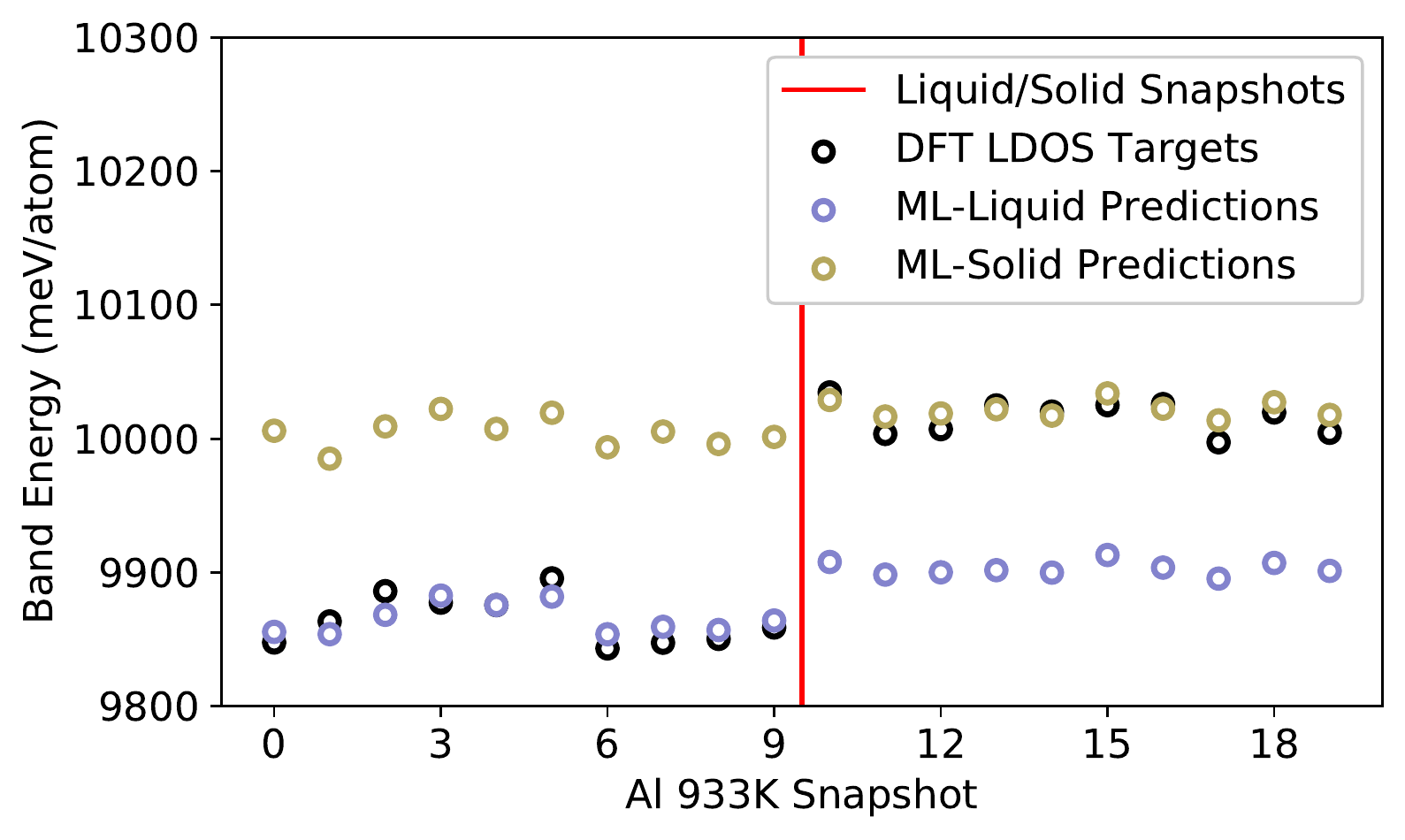}
	\caption{Band energy comparisons for ML-DFT models trained on either 8 liquid or 8 solid snapshots at 933~K. The red line divides the 10 liquid (left) and 10 solid (right) snapshots considered. The ML-Liquid training snapshots are snapshots 0--7 and the validation snapshot is snapshot 8. The ML-Solid training snapshots are snapshots 10--17 and the validation snapshot is snapshot 18.}
	\label{fig:liqsol_bes}
\end{figure}
We demonstrate the errors in total energy and band energy on models trained on data from the same phase, either solid or liquid. Figure~\ref{fig:liqsol_bes} and Table~\ref{tab:errors933k} show these results.
When training, validation, and test sets are all from the same phase, either solid or liquid, the overall ML-DFT accuracy is very high. 
The mean absolute error in the total energy for the 8 snapshot liquid model is 9.31~meV/atom. Moreover, the 8 snapshot solid model has a mean absolute error of 15.71~meV/atom. In both cases, the main contribution to the error is due to the band energy error which is 5.21~meV/atom in the liquid test set and 13.37~meV/atom in the solid. Both models have a similar mean absolute percentage error, 0.01\% in the liquid model and 0.03\% in the solid model.

However, training on a single phase results in poor predictive accuracy for the phase unseen in training, as is shown clearly in Figure~\ref{fig:liqsol_bes} and Table~\ref{tab:errors933k}.
For the ML-DFT model trained on 8 liquid snapshots, the mean absolute total energy error in the solid test set rises to 111.41~meV/atom. For the model trained on 8 solid snapshots, the mean absolute total energy error in the liquid test set rises to 123.29~meV/atom. Again, the major contribution to the error stems from the band energy which is 113.40~meV/atom in the solid test set and 139.96~meV/atom in the liquid.

\subsection{\label{ssec:ml_pred_hybrid-model} Hybrid Liquid-Solid Aluminum ML-DFT Models}
\begin{figure*}[htpb]
	\centering
	\includegraphics[width=\columnwidth]{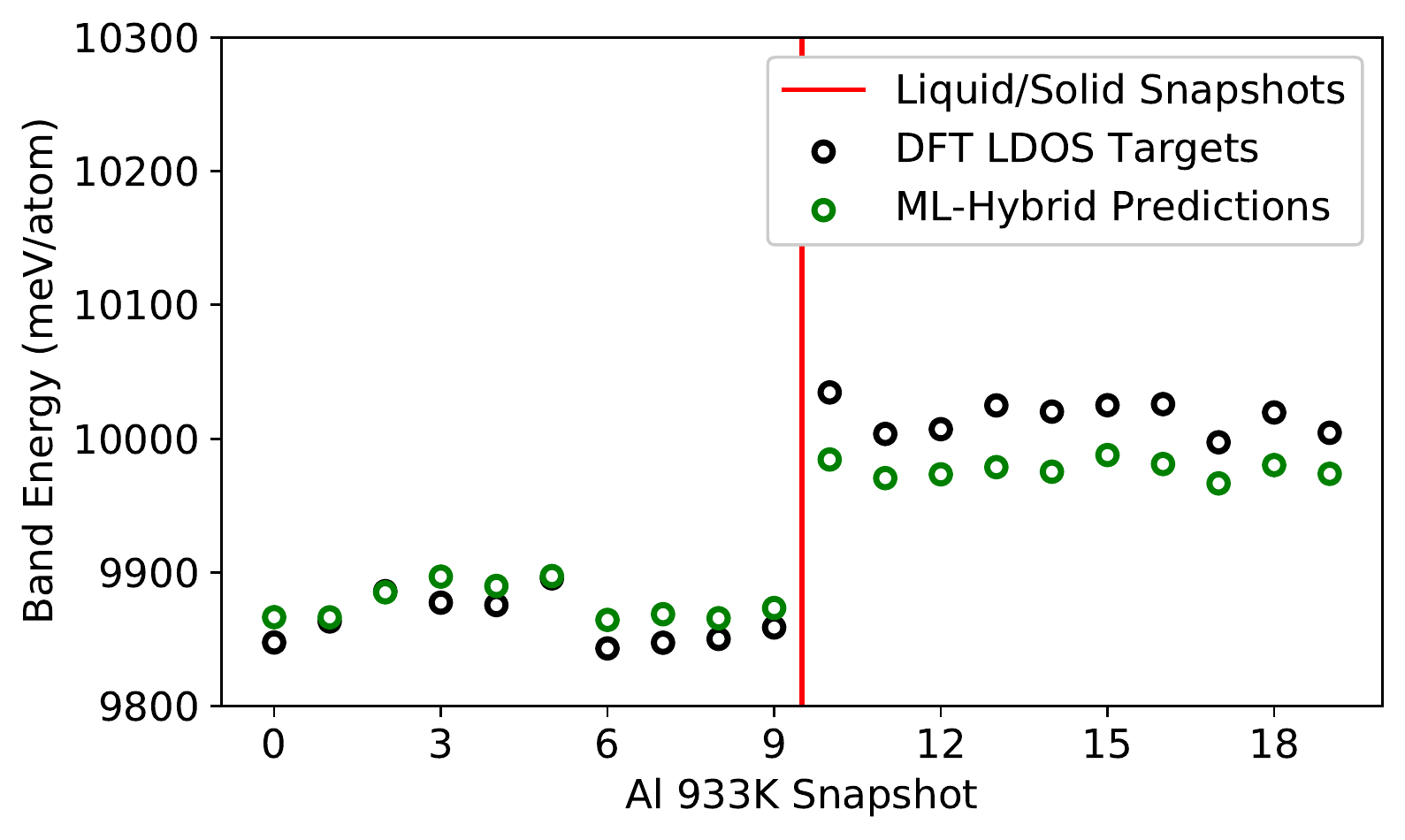}
	\includegraphics[width=\columnwidth]{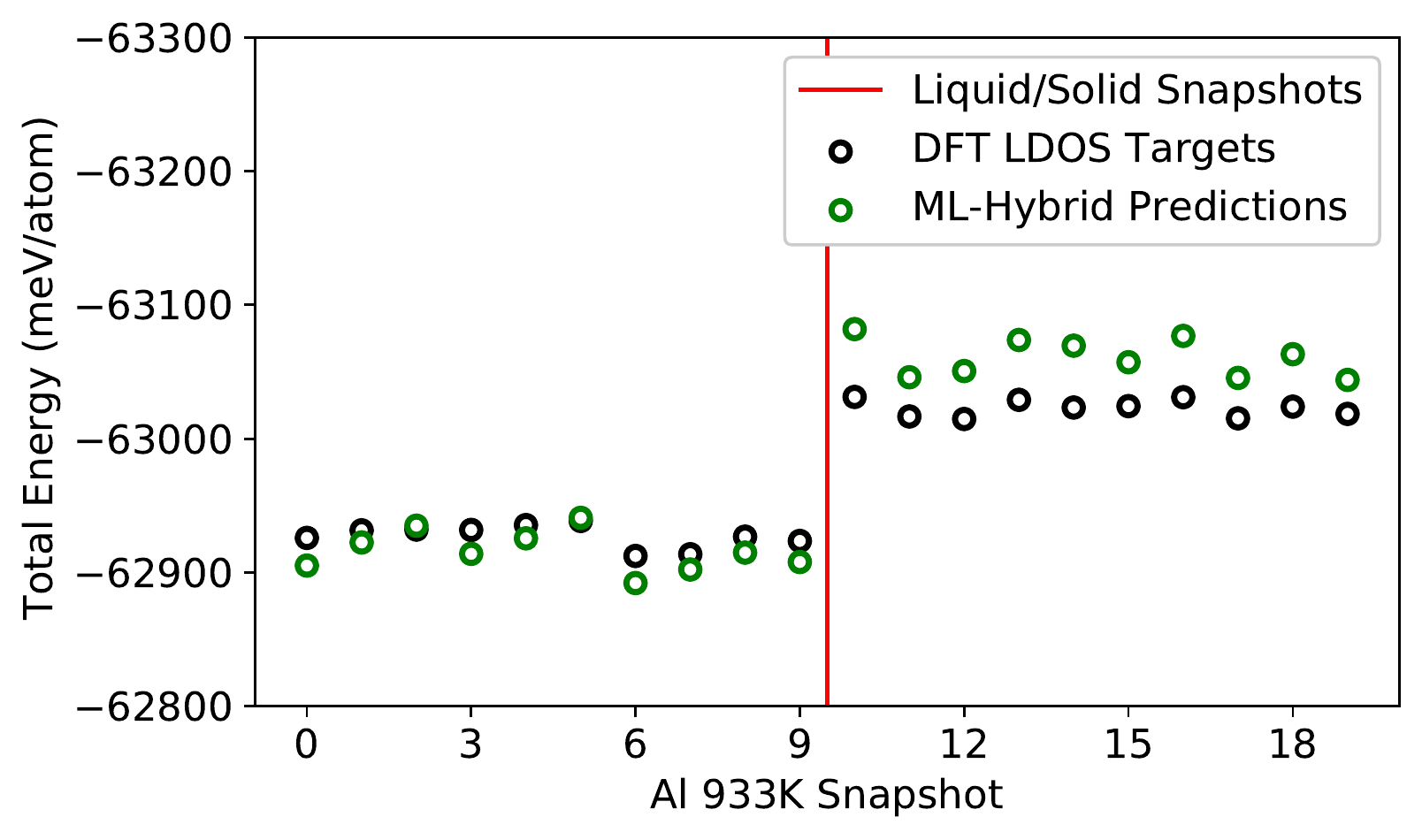}
	\caption{Band energy (left) and total energy (right) comparisons for the ML-DFT model trained on 6 liquid and 6 solid snapshots at 933~K. The red line in each figure divides the 10 liquid (left) and 10 solid (right) snapshots considered. The ML-Hybrid training snapshots are snapshots 0--5 and 10--15 and the validation snapshots are snapshots 6 and 16.}
	\label{fig:hybrid_bes}
\end{figure*}
\begin{table*} [htpb]
	\small
	\begin{center}
		\caption{Band energy and total energy test errors from different ML-DFT models inferred for aluminum snapshots at 933~K. The training set consists of snapshots in liquid or solid states. When trained, validated, and tested 
		on the same phase, either liquid (highlighted in blue) or solid (highlighted in yellow), the overall ML-DFT accuracy is very high.
		The training on a single phase results in poor predictive accuracy for the phase unseen in training, as expected.
		In contrast, our main result, the hybrid ML-DFT model predicts the total energy over both solid and liquid phases with accuracy sufficient for meaningful atomistic simulations (highlighted in green).}
		\begin{ruledtabular}
			\begin{tabular}{ c  c  c  c  c  c  c  c}
				{\bf Training Set}     &  {\bf Test Set} & \multicolumn{3}{c}{\bf Band Energy} & \multicolumn{3}{c}{\bf Total Energy} \\
				{\bf (Validation Set)} & & {\bf Max. AE} & {\bf MAE} & {\bf MAPE} & {\bf Max. AE} & {\bf MAE} & {\bf MAPE} \\
				& & {\bf (meV/atom)} & {\bf (meV/atom)} & {\bf (\%)} & {\bf (meV/atom)} & {\bf (meV/atom)} & {\bf (\%)} \\
				\hline
				4 Liquid             & 5 Liquid  &  13.57 &   9.80 & 0.0994 & 12.39 &  7.13 & 0.0113 \\
				(1 Liquid)           & 10 Solid  & 113.50 & 100.84 & 1.0067 & 106.62 & 91.54 & 0.1452 \\    
		
				\hline
				8 Liquid             & 1 Liquid  &   5.21 &   5.21 & 0.0529 &   \blue{9.31} &   \blue{9.31} & 0.0148\\
				(1 Liquid)           & 10 Solid  & 126.69 & 113.40 & 1.1320 & 125.07 & 111.41 & 0.1768  \\
				
				\hline
				4 Solid              & 10 Liquid & 142.63 & 126.99 & 1.2875 & 142.95 & 126.78 & 0.2015 \\
				(1 Solid)            & 5 Solid   &  13.15 &   8.44 & 0.0843 &  14.45 &   9.81 & 0.0156 \\
				
				\hline
				8 Solid              & 10 Liquid & 158.25 & 139.96 & 1.4190 & 153.45 & 123.29 & 0.1959 \\
				(1 Solid)            & 1 Solid   &  13.37 &  13.37 & 0.1336 &  \yellow{15.71} &  \yellow{15.71} & 0.0249\\

				\hline
				2 Liquid + 2 Solid   & 8 Liquid  &  73.76 &  62.14 & 0.6299 &  57.34 & 48.47 & 0.0770 \\
				(1 Solid)            & 7 Solid   &  37.26 &  28.22 & 0.2816 &  41.53 & 29.50 & 0.0468\\

				\hline
				4 Liquid + 4 Solid   & 6 Liquid  &  67.28 &  58.23 & 0.5906 & 61.29 & 51.37 & 0.0816 \\
				(1 Solid)            & 5 Solid   &  25.33 &  16.16 & 0.1613 & 26.98 & 15.11 & 0.0240 \\				
				\hline
				4 Liquid + 4 Solid   & 5 Liquid  &  36.84 &  28.54 & 0.2896 & 28.39 & 21.33 & 0.0339\\
				(1 Liquid + 1 Solid) & 5 Solid   &  44.94 &  37.19 & 0.3711 & 47.30 & 38.57 & 0.0612\\ 
				\hline		
				6 Liquid + 6 Solid   & 3 Liquid  &  21.34 &  17.11 & 0.1737 &  \green{15.76} & {\bf \green{13.04}} & 0.0207 \\
				(1 Liquid + 1 Solid) & 3 Solid   &  39.33 &  33.56 & 0.3354 & \green{39.23} & {\bf \green{31.60}} & 0.0501\\				
				\hline
			\end{tabular}
		\end{ruledtabular}    
		\vspace{.05in}
		\label{tab:errors933k}
	\end{center}
\end{table*}

In Figure~\ref{fig:hybrid_bes} and Table~\ref{tab:errors933k} we demonstrate our central result, a single ML-DFT model that is capable of yielding accurate results 
for both liquid and solid aluminum at a temperature of 933~K with accuracy sufficient for meaningful atomistic simulations of electronic, thermodynamic, and kinetic behavior of materials.
Using the hyperparameters listed in Table~\ref{tab:ml_params}, the most accurate dual-phase \emph{hybrid} ML-DFT model was trained using 6 liquid and 6 solid snapshots (snapshots 0--5 and 10-15) as the training dataset and 1 solid and 1 liquid snapshot (snapshots 6 and 16) as a validation set. This ML-DFT model achieves state-of-the-art ML accuracy for all test snapshots considered. 

Overall, Table~\ref{tab:errors933k} summarizes the total energy (Eq.~\ref{eq:FES.LDOS}) and band energy (Eq.~\ref{eq:Eb.LDOS}) errors for each of the eight ML-DFT models. Training and validation sets determine the specific ML-DFT model and the test sets are used to measure the model's ability to generalize to liquid or solid phase snapshots. 

Most notably, our overall most accurate hybrid ML-DFT model predicts the total energy over both solid and liquid phases with a mean absolute error of 13.04 meV/atom in the solid test set and 31.60 meV/atom in the liquid test set, as illustrated in Figure~\ref{fig:hybrid_bes} and listed in the last row of Table~\ref{tab:errors933k}. In comparison with the single-phase ML-DFT models in Figure~\ref{fig:liqsol_bes}, the hybrid ML-DFT model is able to generalize to both liquid and solid snapshots. This is adequate to resolve the solid-liquid total energy difference (110~meV/atom) and is approaching the 5~meV/atom accuracy of the best ML-IAPs used in large-scale MD simulations~\cite{Zuo2020}. 

Furthermore, the sensitivity of ML-DFT towards the volume of training data is also assessed in Table~\ref{tab:errors933k}. Reducing the training dataset to only 2 liquid and 2 solid snapshots degrades the accuracy for both the band and total energies as expected. While the mean absolute error of 29.50 meV/atom in the solid phase is about the same as in the larger training set, the error in the liquid phase rises to 48.47 meV/atom. This emphasizes the fact that a desired accuracy in the ML-DFT model can be approached systematically by increasing the volume of training data.

Interestingly, in Figure~\ref{fig:hybrid_bes}, the training snapshots do not necessarily have the lowest error as might be expected. The band and total energy errors for training snapshot 0, for example, are slightly larger than the errors for test snapshots 8 and 9. For the solid snapshots, both training snapshots (10--15) and test snapshots (17--19) have very similar errors in both the band energy and total energy. Similar observations can be made for Figure~\ref{fig:liqsol_bes}. A greater number of liquid and solid test snapshots will only further refine the statistics of the ML-DFT model accuracy. Future investigations will include both sensitivity analysis and thorough validation of the ML-DFT model, particularly in the context of large-scale MD simulations. 

\section{\label{sec:conclusion} Discussion}
In the present work, we establish an ML framework that eliminates the most computationally demanding and poorly scaling components of DFT calculations.  Once trained with DFT results for similar systems, it can predict many of the key results of a DFT calculation, including the LDOS, DOS, density, band energy, and total energy, with good accuracy. After training, the only input required by the model is the atomic configuration, i.e., the positions and chemical identities of atoms. We have been able to train a single model that accurately represents both solid and liquid properties of aluminum. The ML-DFT model is distinct from previous ML methodologies, as it is centered on materials modeling in support of accurate MD-based predictions for the evolution of both electronic and atomic properties. 
We represent materials descriptors (such as the LDOS) on a regular grid as opposed to basis functions~\cite{BVL+2017:bypassing} and use a specific formulation in terms of the LDOS to reconstruct the total energy.  Our approach also generalizes to non-zero electronic temperatures in a straightforward way, which allows it to be easily applied to materials at extreme conditions.

Furthermore, the very recently published work~\cite{CKB+2019:solving} stopped short of demonstrating that the ML-predicted LDOS can be utilized to accurately calculate energies and forces needed to enable MD calculations.
Further work on computing accurate forces of materials under elevated temperatures and pressures will follow, as well as a detailed exploration of ML-DFT's accuracy when scaled to O($10^4$) atoms, well beyond the reach of standard DFT.

\section*{Acknowledgments}

Sandia National Laboratories is a multimission laboratory managed and operated by National Technology \& Engineering Solutions of Sandia, LLC, a wholly owned subsidiary of Honeywell International Inc., for the U.S. Department of Energy’s National Nuclear Security Administration under contract DE-NA0003525. This paper describes objective technical results and analysis. Any subjective views or opinions that might be expressed in the paper do not necessarily represent the views of the U.S. Department of Energy or the United States Government. This work was partly funded by the Center for Advanced Systems Understanding (CASUS) which is financed by the German Federal Ministry of Education and Research (BMBF) and by the Saxon State Ministry for Science, Art, and Tourism (SMWK) with tax funds on the basis of the budget approved by the Saxon State Parliament. We thank Warren Davis and Joshua Rackers for helpful conversations and Mitchell Wood for visualizing the SNAP fingerprint grid.

\appendix

\section{\label{app:ldos-eval} Analytical Evaluation of Derived Quantities from the Local Density of States}

Consider the numerically challenging integral in Eq.~(\ref{eq:form_of_integrals}). When $g = f^\beta(\epsilon)$, this integral gives the electronic density and the total number of electrons, when $g = \epsilon f^\beta(\epsilon)$ the result is the band energy, and when $g = g^S \equiv \beta^{-1} \left\{ f^\beta(\epsilon) \log[f^\beta(\epsilon)] + [1-f^\beta(\epsilon)] \log[1-f^\beta(\epsilon)] \right\}$ we obtain the entropy contribution to the energy. When $D(\epsilon) = \bar{D}[D](\epsilon;\ubR)$, Eq.~(\ref{eq:form_of_integrals}) results in a scalar quantity, and we can evaluate the total number of electrons, the band energy, or the entropy contribution to the total energy.  Alternatively, when $D(\epsilon) = D(\epsilon,\br; \ubR)$, Eq.~(\ref{eq:form_of_integrals}) results in a field that depends on the position $\br$ with the electronic density being the most important result.

Our ML model gives us $D(\epsilon_i)$ evaluated on a grid of energy values $\epsilon_i$.  We will extend $D(\epsilon_i)$ to a $D(\epsilon)$ defined for all $\epsilon$ between $\epsilon_0$ and $\epsilon_N$ by linear interpolation.  This allows us to perform the required integrals analytically.  It is necessary to treat these integrals carefully because $f^\beta(\epsilon)$ changes rapidly compared to $D(\epsilon)$ for many systems of interest.  In particular, $f^\beta(\epsilon)$ changes from one to zero over a few $k_BT$, while the spacing of the $\epsilon_i$ grid is several times $k_BT$ at room temperature.  In order to evaluate these integrals analytically, we can use the representation of $f^\beta(\epsilon)$ and $g^S(\epsilon)$ in terms of the polylogarithm special functions $\operatorname{Li_n}(x)$
\begin{align}
	f^\beta(x) &= 1 + \operatorname{Li_0}(-e^x) \\
	g^S(x) &= \beta^{-1} \left[-x\operatorname{Li_0}(-e^x) + \operatorname{Li_1}(-e^x)\right]
\end{align}
where $x = \beta (\epsilon - \mu)$.  It is useful to define a series of integrals giving moments of $f^\beta$ and $g^S$ with respect to $\epsilon - \mu$ and evaluate these integrals using integration by parts and the properties of $\operatorname{Li_n}(x)$.  In particular, the relationship
\begin{equation}
	\frac{d\operatorname{Li_n}(-e^x)}{dx} = \operatorname{Li_{n-1}}(-e^x)
\end{equation}
is useful.  The required integrals are

\begin{align}
	\begin{split}
		F_0 &\equiv \beta^{-1} \int dx\ f^\beta(x)  \\
		&= \beta^{-1} \left[ x + \operatorname{Li_1}(-e^x) \right]\,,
	\end{split}\\
	\begin{split}
		F_1 &\equiv \beta^{-2} \int dx\ x\, f(x)  \\
		&= \beta^{-2} \left[ \tfrac{x^2}{2}  
		+ x \operatorname{Li_1}(-e^x)
		- \operatorname{Li_2}(-e^x) \right]\,,
	\end{split}\\
	\begin{split}
		F_2 & \equiv \beta^{-3} \int dx\ x^2 f(x)  \\
		& = \beta^{-3} \left[ \tfrac{x^3}{3}  
		+ x^2 \operatorname{Li_1}(-e^x)
		- 2x \operatorname{Li_2}(-e^x) \right. \\
		&\quad \left. + 2 \operatorname{Li_3}(-e^x) \right]\,,
	\end{split}\\
	\begin{split}
		S_0 &\equiv \beta^{-1} \int dx\ g^S(x)  \\
		&=  \beta^{-2} \left[-x\operatorname{Li_1}(-e^x)
		+ 2\operatorname{Li_2}(-e^x)\right]\,,    
	\end{split}\\
	\begin{split}
		S_1 &\equiv \beta^{-2} \int dx\ x g^S(x) \\
		&=  \beta^{-3} \left[-x^2\operatorname{Li_1}(-e^x)
		+ 3x\operatorname{Li_2}(-e^x)  \right.\\
		&\quad \left. - 3\operatorname{Li_3}(-e^x)\right].    
	\end{split}
\end{align}

Quantities of interest with the form of Eq.~(\ref{eq:form_of_integrals}) can then be evaluated as a weighted sum over the DOS or LDOS evaluated at the energy grid points
\begin{equation}
	I = \sum_i w_i D(\epsilon_i),
\end{equation}
where the weights $w_i$ are given by
\begin{align}
	\begin{split}
		w_i &= \left[I_0(\epsilon_{i+1}) - I_0(\epsilon_{i})\right] 
		\left[ 1 + \frac{\epsilon_i - \mu}{\epsilon_{i+1} - \epsilon_{i}}\right] \\
		&\quad + \left[I_0(\epsilon_{i}) - I_0(\epsilon_{i-1})\right] 
		\left[1 - \frac{\epsilon_i - \mu}{\epsilon_{i} - \epsilon_{i-1}}\right]  \\
		&\quad - \frac{I_1(\epsilon_{i+1}) - I_1(\epsilon_{i})}{\epsilon_{i+1} - \epsilon_{i}}
		+ \frac{I_1(\epsilon_{i}) - I_1(\epsilon_{i-1})}{\epsilon_{i} - \epsilon_{i-1}}\ .
	\end{split}
\end{align}
When $D(\epsilon) = \bar{D}[D](\epsilon;\ubR)$, $I_0 = F_0$ and $I_1 = F_1$, we obtain $I = N_e$, the total number of electrons in the system.  The Fermi level $\mu$ is then adjusted until $N_e$ is equal to the total number of valence electrons in the system (i.e., the system is charge neutral).  When $I_0 = F_1$ and $I_1 = F_2$, we obtain $I = E_{b}[D] - \mu N_e$, and we can easily obtain the band energy $E_b[D]$ by adding $\mu N_e$.  When $I_0 = S_0$ and $I_1 = S_1$, we obtain $I = -\beta^{-1}S_S[D]$, the entropy contribution to the energy.  Finally, when $D(\epsilon) = D(\epsilon,r; \ubR)$, $I_0 = F_0$ and $I_1 = F_1$, we obtain $n[D](\br)$, the electronic density.  Given these quantities, it is straightforward to evaluate the total (free) energy using Eq.~(\ref{eq:FES.LDOS}) and the standard methods of DFT.

\section{\label{app:ml-model-details} Machine-learning Model Details}

\begin{table} [htpb]
	\small
	\begin{center}
		\caption{Optimized hyper-parameters for the 298K and 933K models.}
		\begin{ruledtabular}
			\begin{tabular}{ c  c  c }
				{\bf Model}     & {\bf 298K}  & {\bf 933K}      \\
				{\bf Parameter} & {\bf Model} & {\bf Models}    \\
				\hline
				FP length         & 91            & 91        \\
				FP scaling        & Elem.-wise stand. & Elem.-wise stand. \\
				LDOS length       & 250           & 250       \\
				LDOS scaling      & Max norm. & Max norm. \\
				\hline
				Optimizer         & Adam          & Adam      \\
				Minibatch size    & 1000          & 1000      \\
				Learning rate (LR)& 1e-5          & 5e-5      \\
				LR schedule       & 4 epochs   & 4 epochs  \\
				Early stopping    & 8 epochs & 8 epochs \\
				\hline
				Max layer width   & 800           & 4000      \\
				Layers            & 5             & 5         \\
				Activation Func.  & LeakyReLU     & LeakyReLU \\
				Total weights     & 1.62e6        & 3.34e7    \\
			\end{tabular}
		\end{ruledtabular}
		\label{tab:ml_params}
	\end{center}
\end{table}

\begin{figure}[ht!]
	\centering
	\vspace{0.5cm}
	\includegraphics[scale=0.8]{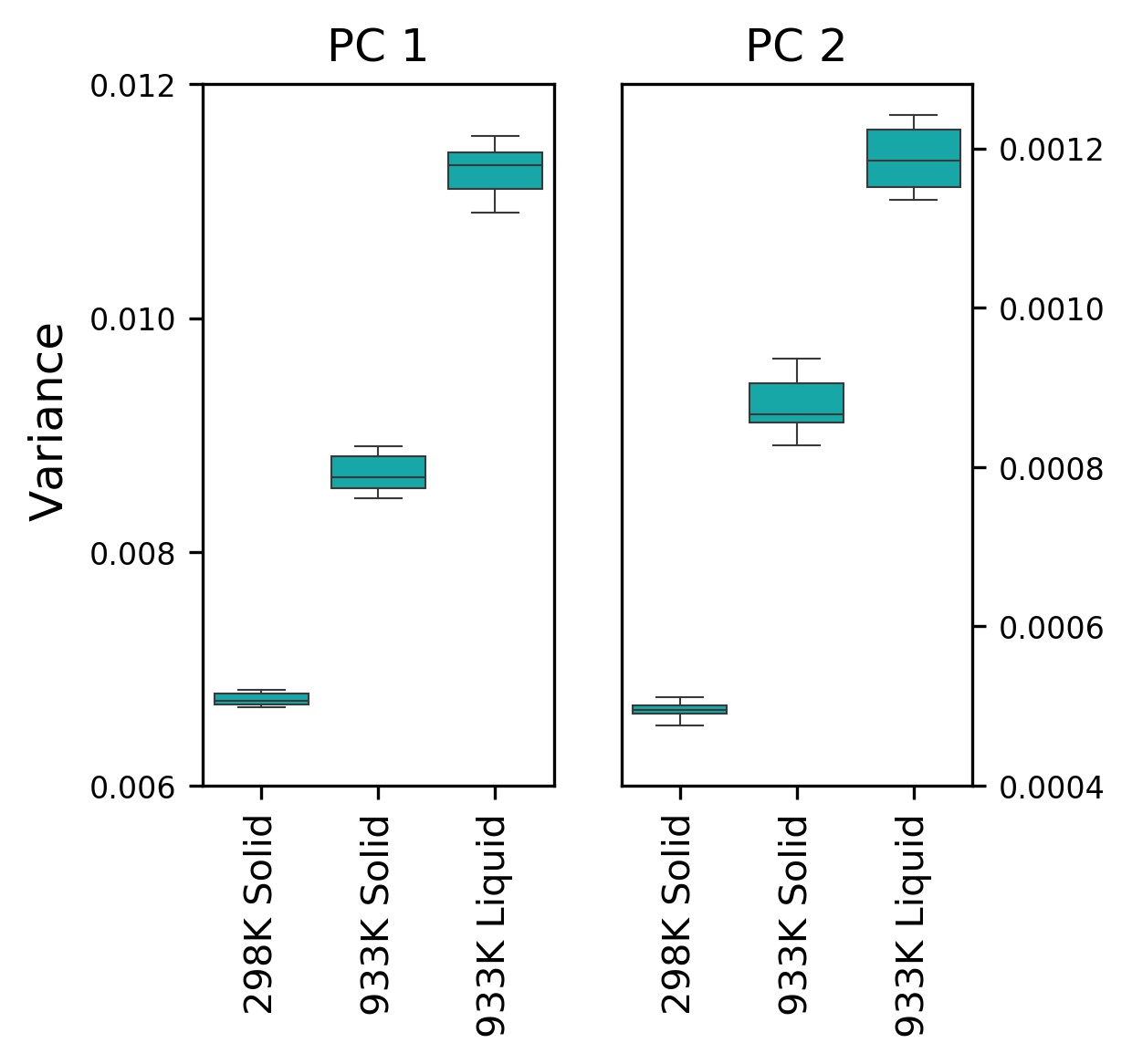}
	\caption{Variance of the first and second principal component scores for each group of snapshots, 298~K solids, 933~K solids, and 933~K liquids. Each group contains 10 snapshots. The boxes enclose the second and third quartiles, and the whiskers indicate the minimum and maximum values.}
	\label{fig:ldos_var}
\end{figure}

The 298~K systems include 10 snapshots of SNAP fingerprint and Quantum Espresso LDOS data. At 933~K, there are 10 snapshots of liquid aluminum and 10 snapshots of solid aluminum. For the fingerprints and LDOS data, each snapshot uses the same uniform Cartesian real-space grid, ($ 200 \times 200 \times 200 = 8 \times 10^6 $ points). Each snapshot, including both SNAP fingerprint and LDOS data, is 20.7~gigabytes.

We observe that the 933~K ML-DFT models require a greater amount of training data. Furthermore, the optimized 298~K model requires approximately 1.6 million weights, whereas the 933~K models require 33.4 million weights. The larger training set and greater model complexity can be attributed to greater LDOS variability at the higher temperature.

To facilitate comparing LDOS variability between the three groups (298~K solid, 933~K solid, and 933~K liquid), the dimensionality of the LDOS datasets was reduced using principal component analysis (PCA). Because of the large amount of data ($ 30 \times 8 \times 10^6 $ samples by 250 energy levels), PCA was performed incrementally using the scikit-learn Python package~\cite{scikit-learn}. The resulting first principal component explained approximately 81.7\% of the variance in the LDOS, and the second explained an additional 7.9\%.

The variance of the first two principal component scores was calculated for each of the 30 snapshots. Figure~\ref{fig:ldos_var} shows these variances grouped by snapshot temperature and phase. 933~K liquid snapshots are seen to possess the highest variance, followed by the 933~K solid snapshots. The 298~K solid snapshots have the lowest variance.

\begin{table*} [t!]
	\small
	\begin{center}
		\caption{Band energy and total energy test errors from the ML-DFT models inferred for aluminum snapshots at 298~K.}
		\begin{ruledtabular}
			\begin{tabular}{ c  c  c  c  c  c }
				{\bf Training Set}     &  {\bf Test Set} & \multicolumn{2}{c}{\bf Band Energy} & \multicolumn{2}{c}{\bf Total Energy} \\
				{\bf (Validation Set)} & & {\bf Max. AE} & {\bf MAE} & {\bf Max. AE} & {\bf MAE} \\
				& & {\bf (meV/atom)} & {\bf (meV/atom)} & {\bf (meV/atom)} & {\bf (meV/atom)} \\
				\hline
				1 Solid   & 8 Solid & 3.82 & 2.48 & 5.60 &  {\bf 3.07} \\
				(1 Solid) &         &      &      &      &      \\    
			\end{tabular}
		\end{ruledtabular}
		\vspace{.05in}
		\label{tab:errors298k}
	\end{center}
\end{table*}

\begin{figure*}[htpb]
	\centering
	\includegraphics[width=\columnwidth]{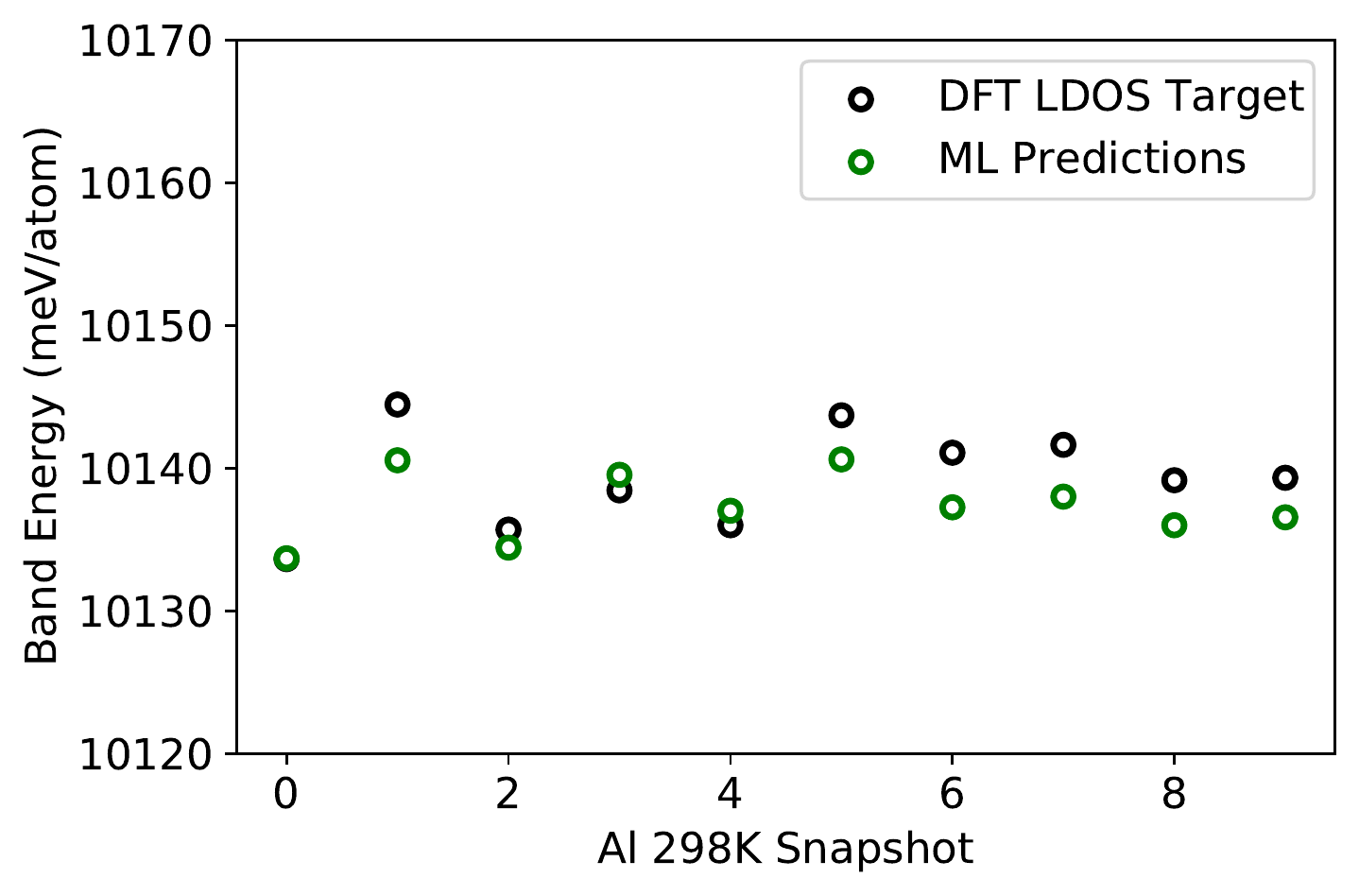}
	\includegraphics[width=\columnwidth]{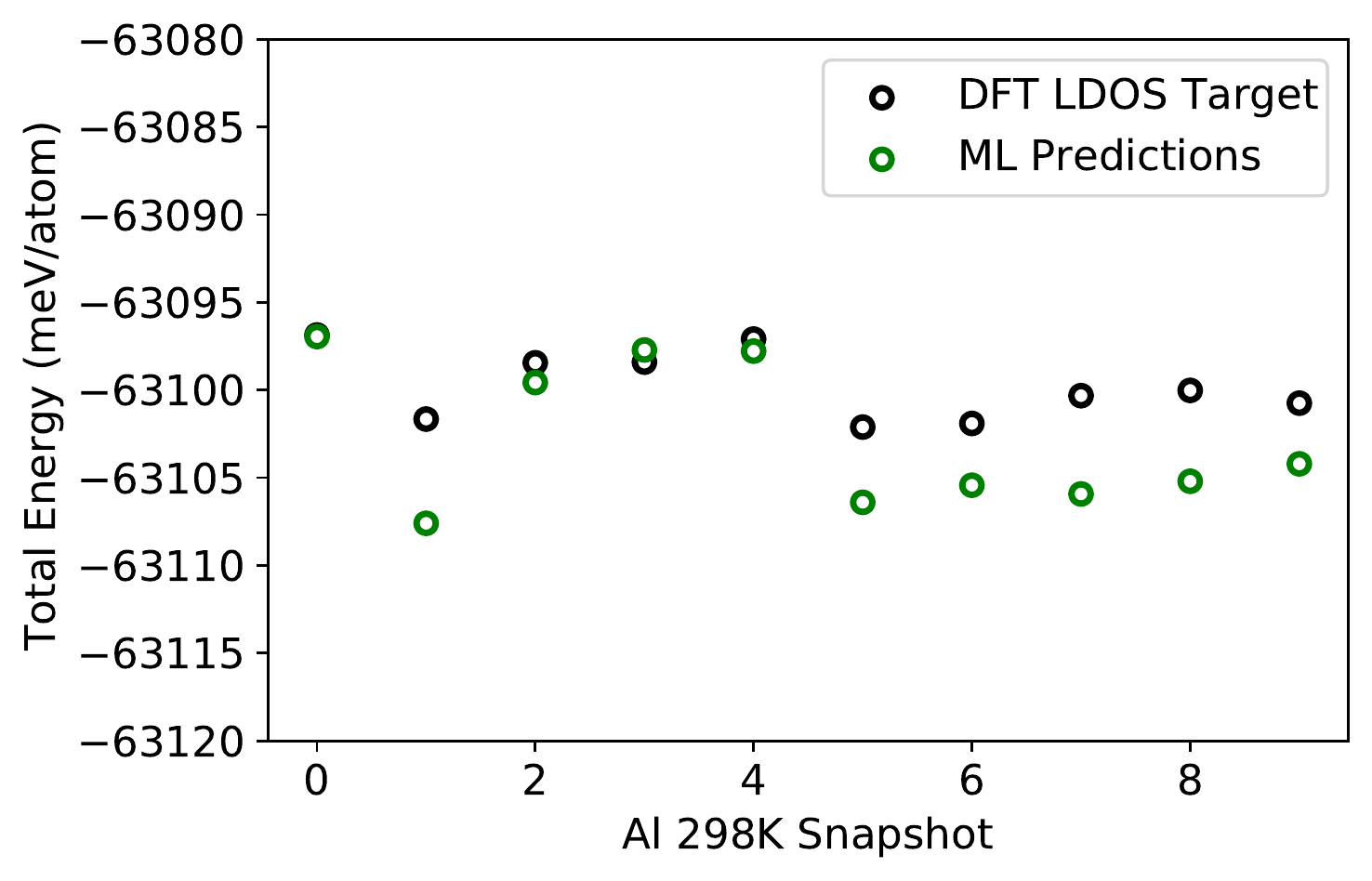}
	\caption{Band energy (left) and total energy (right) comparisons for the ML-DFT model trained on 1 solid snapshot at 298~K. The training and validation snapshot are snapshot 0 and 1, respectively.}
	\label{fig:solid298k}
\end{figure*}
\begin{figure*}[htpb]
	\centering
	\includegraphics[width=\textwidth]{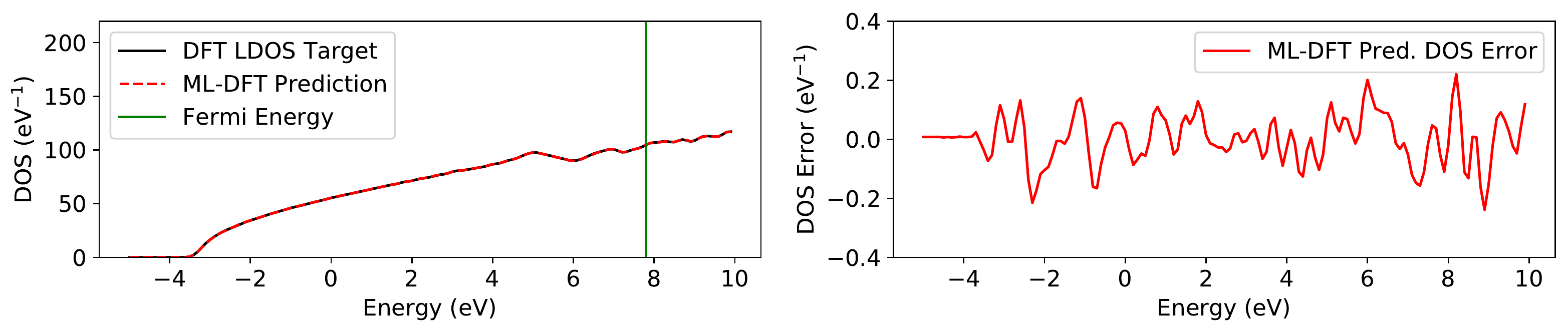}
    \caption{DOS comparison for a test snapshot using the ML-DFT model trained on snapshots at 298~K. The vertical line indicates the Fermi energy which is 7.797~eV.}
	\label{fig:solid298k.DOS}
\end{figure*}

\begin{table*} [t!]
	\small
	\begin{center}
		\caption{Timings in different steps contributing to the total evaluation of KS-DFT quantities based on neural network inference. The reported times are averaged over the inference of 10 snapshots.}
		\begin{ruledtabular}
			\begin{tabular}{ p{0.3\textwidth} p{0.1\textwidth}  p{0.1\textwidth} p{0.4\textwidth} }

				{\bf Step}     & {\bf Time (s) }   & {\bf Time (s) } & {\bf Optimization notes}      \\
				            &    {\bf 298K Model}    &      {\bf 933K Model}       & \\
				\hline
				Generate fingerprint vectors from atomic positions         & 2909.75 & 2908.12            & LAMMPS used only in serial mode, speedup possible by using parallelization        \\
				Load modules        & 0.36 & 0.36 & -- \\
				{\bf Infer LDOS from neural network}     & {\bf 18.72}  & {\bf 53.94}           & {\bf Only a single GPU was used}        \\
				Integrate LDOS to DOS & 1.57     & 1.44 & -- \\
				Determine Fermi level    & 2.23  & 2.40 & -- \\
				Integrate LDOS to Density  & 1282.39    & 1277.66 & Integration  unoptimized and in serial \\
				Evaluate total energy contributions from DOS & 0.96 & 0.99 & -- \\
				Evaluate total energy contributions from density & 87.50 & 87.17 & Quantum Espresso used only in serial mode, speedup possible by using parallelization  \\
				
			\end{tabular}
		\end{ruledtabular}
		\label{tab:inference_timing}
	\end{center}
\end{table*}

We perform a sequence of hyperparameter tunings once for each temperature using the optimization strategy described in Section~\ref{ssec:ml_workflow}. The final 933~K models are trained using the same hyperparameters. Further accuracy may be obtainable once training throughput is increased and multiple, independent hyperparameter optimizations are made possible. The optimized ML-DFT model parameters are given in Table~\ref{tab:ml_params}.
Scaling of the input and output data prior to training is critical to ML-DFT model accuracy. We perform an \emph{element-wise} standardization to mean 0 and standard deviation 1 for the fingerprint inputs. For example, standardize the first scalar of all fingerprint vectors in the training set, then standardize the second scalar of all fingerprints, and so on. For the LDOS outputs, we perform a normalization using the maximum LDOS value of the entire training dataset. 

The hyperparameter convergence for the 298~K and 933~K models required 48 and 23 training iterations, respectively. Once the 298~K case was optimized, using those optimized hyperparameters as the 933~K optimization's initial iterate greatly accelerated convergence resulting in fewer training iterations. Within an iteration, the 298~K model requires 93~epochs. For the 8 snapshot 933~K cases, the single phase models, either liquid or solid, required just 39--50~epochs to converge. Comparatively, the final hybrid liquid-solid model required 101 epochs to converge. 

Due to the large quantity of training data in the 933~K case (12 training $+$ 2 validation snapshot $\times$ 20.7~gigabytes), the required training strategy for single node machines is to \emph{lazily load} the snapshots into memory each epoch. This strategy is needed for machines where the CPU memory is insufficient and allows ML-DFT to train on any arbitrary number of snapshots. All ML-DFT models have been trained on a single node with 1 NVIDIA Tesla GPU. The 298~K network, with only 1 training and 1 validation snapshot, does not need to lazily load snapshots, and so each epoch requires approximately 151 seconds. For the 933~K networks, the total runtime of a single epoch with 12 training snapshots is approximately 76 minutes. With a sufficient number of nodes, we are able to stage each training snapshot on an independent node to alleviate the lazy loading data movement penalty completely. Future performance improvements to the workflow will target this bottleneck chiefly. 

Finally, ML-DFT inference requires only 54 seconds to obtain a full grid of LDOS predictions for one snapshot. A full overview of the computational time required for each individual step of inference is given in Tab.~\ref{tab:inference_timing}. Apart from neural network inference, the major time contributions to overall inference time stem from the calculation of fingerprint vectors, LDOS integration and evaluation of density dependent terms. Parallelization and/or optimization is possible for these calculation steps, with drastic speedups to be expected.
Once the cost of creating a ML-DFT model is sufficiently amortized, the computational speed up relative to standard DFT becomes quite stark.

\section{\label{app:Al-solid-298K} Solid Aluminum at Ambient Temperature and Pressure}

We consider modeling of the ambient temperature and pressure aluminum systems as a base case for all future work and experiments. Future investigations of the grid-based ML-DFT model, such as an extensive exploration of alternative ML models, fingerprints, LDOS formulations, and non-aluminum systems, should use this baseline in terms of ML accuracy and computational performance. 

In Table~\ref{tab:errors298k} and Figure~\ref{fig:solid298k}, we demonstrate that the 298~K model is able to achieve the highest and most consistent accuracy of all models considered in this work. 
Using a single training snapshot and a single validation snapshot, the ML-DFT model is able to achieve mean absolute band energy error of 2.48~meV/atom and mean absolute total energy error of 3.07~meV/atom. 
Furthermore, the 298~K model surpasses the 5~meV/atom accuracy for the most accurate ML-IAPs in large-scale MD simulations~\cite{Zuo2020}. 
The low variability, seen in Figure~\ref{fig:ldos_var}, is more easily exploited by ML-DFT's grid-based approach relative to the 933~K cases. 


%


%

\end{document}